\documentclass[
  aps,
  prd,
  reprint,            % PRD two-column look; switch to 'preprint' for 1-col drafts
  superscriptaddress,
  nofootinbib,
  longbibliography     % show titles only if your .bst does; keeps refs tidy
]{revtex4-2}

\usepackage[T1]{fontenc}
\usepackage[utf8]{inputenc}
\usepackage{lmodern}
\usepackage{microtype}
\usepackage{subfigure}

\usepackage{amsmath,amssymb,amsfonts}
\usepackage{bm}        % bold math (vectors, tensors)
\usepackage{mathtools} % fixes & extras for amsmath
\usepackage{physics}   % optional: \dv, \pdv, \qty, etc.
\usepackage{float}
\usepackage{tikz}
\usetikzlibrary{snakes}
\usetikzlibrary{decorations.pathmorphing}
\usepackage{graphicx}
\usepackage{dcolumn}   % align table columns on decimal point
\usepackage{xcolor}
\usepackage{tikz}
\usetikzlibrary{arrows.meta,calc,decorations.pathmorphing}

\usepackage[colorlinks=true,linkcolor=blue,citecolor=blue,urlcolor=blue]{hyperref}

\usepackage[capitalise]{cleveref}

\renewcommand{\vec}[1]{\bm{#1}}

\tikzset{
    graviton/.style={
        decorate,
        decoration={coil, aspect=0, segment length=5pt, amplitude=1pt},
        double,
        double distance=0.5pt
    }
}
\begin{document}

\title{Graviton Photoproduction by a Kerr--Newman Black Hole with Worldline EFT}

\author{Qinyuan Zheng}
\email{qinyuan.zheng@yale.edu}
\affiliation{Department of Physics, Yale University, New Haven, CT 06511, USA}

\date{\today} % or a fixed date; add \preprint{arXiv:NNNN.NNNNN} if desired

% --- Abstract must come before \maketitle in REVTeX ---
\begin{abstract}
We present the first computation of the gauge-invariant, long-wavelength scattering amplitude for the graviton photoproduction by a Kerr--Newman black hole through $\mathcal{O}\big((\omega m)^2\big)$, or correspondingly $\mathcal{O}(S^2)$, and to linear order in $G$, using the worldline effective field theory. We show that electromagnetic interactions can be introduced consistently into the spinning worldline theory while preserving spin gauge invariance. We also derive the full angular dependence of the conversion cross section through $\mathcal{O}(S^2)$, and demonstrate that the relevant Wilson coefficients at this order are fixed entirely by matching the electromagnetic and gravitational multipole moments to the Kerr--Newman solution. This result provides a benchmark for future analyses of coupled gravitoelectromagnetic scattering in spinning, charged compact-object backgrounds.
\end{abstract}

\maketitle

% --- Main text ---
\section{Introduction}
\label{sec:intro}
Understanding the interplay between electromagnetism (EM) and gravity is of fundamental importance in both field theoretical and astrophysical/cosmological contexts. While electromagnetic waves and gravitational waves (GWs) propagate largely independently in vacuum, the couplings between them are generically expected in the presence of strong fields. For example, the process in which electromagnetic waves are converted to gravitational waves and vice versa in strong uniform magnetic fields is known as the Gertsenshtein-Zel'dovich effect~\cite{Gertsenshtein1962, Zeldovich1973}. In a quantum field theoretic framework, the process can be described as a graviton-to-photon conversion. This conversion in non-uniform EM fields has also been studied, and may play a non-negligible role in environments where strong electromagnetic fields and strong gravity coexist, such as in the vicinity of compact objects like neutron stars and black holes~\cite{McDonald:2024nxj, Matsuo:2025blj, Ito:2023fcr, Aly:2025dbx}. Moreover, the graviton-photon mixing could also be significant in the early Universe under extreme conditions~\cite{Linet1990, MarklundBrodinDunsby2000, DolgovEjlli2012, Tseng:2025fjf}, potentially leaving an imprint on the cosmic microwave background (CMB)~\cite{Domcke:2020yzq, Lella:2024dus}. 

Among others, the EM-gravity conversion by an electromagnetically charged particle is of fundamental importance, as such a model applies across scales from elementary particles to macroscopic objects. The conversion of EM waves by the Coulomb field of a fixed electric charge in the low energy regime has been calculated using Feynman perturbative techniques~\cite{PhysRevD.16.2915, PhysRevD.91.064008}. A macroscopic realization of such scenarios is the Reissner-Nordstrom (RN) black hole of mass $m$ and charge $e$, which has been studied extensively over the years~\cite{PhysRevLett.31.1317, JohnstonRuffiniPeterson1974, JohnstonRuffiniZerilli1974, Zerilli1974, Gerlach1974, Sibgatullin1974, OlsonUnruh1974, Moncrief1974, Moncrief1975, ChitrePriceSandberg1975, Boughn1975, Matzner1976, Chandrasekhar1979, Gunter1980, BreuerRosenbaumRyanMatzner1981, Hadj_2022}. However, a general separation of variables necessary to solve the Teukolsky equations in the black hole perturbation theory (BHPT) has not been obtained when spin is introduced. This is because the angular and radial gravito-electromagnetic eigenfunctions do not decouple in general for the Kerr-Newman (KN) black holes~\cite{Pani:2013ija, PhysRevLett.38.1505, DudleyFinley1979, BellezzaFerrari1984}. Alternative techniques with certain assumptions have been developed in the slow-rotation limit to study the gravitoelectromagnetic perturbations of spinning black holes~\cite{Pani:2013ija, Deng:2025atg}. In these analyses, the spin parameter is taken to be small and the Einstein-Maxwell equations are linearized with respect to both the oscillation amplitude and the spin parameter. 

It is of both theoretical interest and potential phenomenological significance to study the graviton-photon mixing around classical spinning charged particles. Despite the rich literature on graviton-photon mixing in external fields and in cosmological contexts, the corresponding process mediated by compact charged spinning objects has not been systematically analyzed at the level of classical scattering amplitudes. Previous amplitude calculations have focused on charged scalars~\cite{Ahmadiniaz_2020, Holstein_1_2006} or quantum spinning fields with spin-1/2 and spin-1~\cite{Holstein_1_2006, PhysRevD.91.064008}. Yet, such a treatment is both timely and useful: it connects modern amplitude methods in curved spacetime with astrophysical observables, clarifies the hierarchy of operators responsible for electromagnetic–gravitational conversion, and provides a benchmark against which future solutions to the coupled Teukolsky equations can be compared. In addition, it will be another testbed for studying the relationship between classical and quantum amplitudes, with two gauge sectors of nature simultaneously present.

From the perspective of worldline effective field theory (EFT), the dynamics of compact objects interacting with long-wavelength fields can be systematically organized in terms of a tower of worldline operators~\cite{PhysRevD.73.104029, Goldberger:2006bd, Porto_2006, Levi_2015, Cheung_2018, Cheung_2020, Bern_2021}, with the compact object treated as a point particle along the worldline viewed from afar. The flat space-time trajectories are treated as background fields for the worldline degrees of freedom, and act as sources of worldline and external field fluctuations. Since the discovery of GWs~\cite{LIGO2016, LIGO2017GW170817}, precision theoretical calculations are increasingly important for generating the GW waveforms from binary systems. The worldline EFT sees success in such calculations in the perturbative ``inspiral'' phase, often organized as the post-Newtonian (PN) approximation~\cite{PhysRevD.73.104029, Goldberger:2006bd, Porto_2006, Levi_2015}, where expansion is performed on gravitational constant $G$ and velocity $v$, or the post-Minkowski (PM) approximation~\cite{Cheung_2018, Cheung_2020, Bern_2021, Cheung_2020, cangemi2023kerrblackholesmassive, Damgaard_2019, Chen_2022, Chung_2019}, where expansion is performed on $G$ only. Apart from calculating Hamiltonians and radiation~\cite{Jakobsen_2022, Jakobsen_2023, Jakobsen:2022psy, Jakobsen:2023ndj, Jakobsen:2023hig}, worldline EFT has also seen its use in bound observables~\cite{Kalin:2019inp, Kalin:2019rwq, Vaidya_2015, Neill_2013, Damour_2018, Bjerrum_Bohr_2020}. Recent interest in the relationship between quantum and classical scattering amplitudes has yielded useful insights, including classical limits in amplitudes and classical observables~\cite{Kosower:2018adc, Maybee:2019jus, Cristofoli:2021vyo}, classical double copy~\cite{BernCarrascoJohansson2010, MonteiroOConnellWhite2014, Luna2015, GoldbergerLi2017, Shen2018, Guevara2019, Lee:2018gxc, White2021Review, Bern:2019prr}, and classical spin effects in particular~\cite{Porto_2006, Porto:2010tr, Levi_2015, PhysRevD.106.124026, benshahar2023scatteringspinningcompactobjects}. 

In this work, we set up the worldline EFT for graviton-photon mixing in the presence of a spinning charged compact object based on the framework developed in~\cite{benshahar2023scatteringspinningcompactobjects, steinhoff2015spingaugesymmetryaction}, and provide the first explicit calculation of graviton-photon conversion amplitude by scattering with a KN black hole. Unlike in BHPT where the full curvature and field solution is attempted near the horizon, we work in the long-wavelength regime, where the compact object can be modeled as a classical point particle with spin-induced multipole moments, and include the non-minimal operators up to $\mathcal{O}(S^2)$. We observe that at this order, the scattering amplitude depends only on the long-distance multipole structure of the background, allowing one to compute the conservative mixed graviton-photon response without solving the coupled perturbation equations. We derive the tree-level graviton to photon scattering amplitude, check its gauge invariance, and discuss the role of spin couplings in the scattering cross section. Finally, we discuss the implications of our results and prospects of future investigations. We keep $c=\hbar=1$ throughout unless otherwise stated, and use mostly minus convention for the metric tensor.

The paper is organized as follows: in Sec.~\ref{sec:setup} we set up the worldline EFT, in Sec.~\ref{sec:matching} we match the theory to the KN black hole, in Sec.~\ref{sec:feynmanrules} we derive the Feynman rules, in Sec.~\ref{sec:results} we present our results, and in Sec.~\ref{sec:discussion} we discuss the implications and future directions.

\section{EFT of spinning charged particle}
\label{sec:setup}
A compact object, such as a black hole or a neutron star, can be treated as a point particle worldline defect in the long-wavelength regime ($\lambda\gg r_s$), where $\lambda$ is the wavelength of the electromagnetic and gravitational waves, $r_s$ is the characteristic size of the particle---usually taken to be the horizon radius for a black hole. In this limit, the internal structure or local curvature structure near the compact object can be integrated out and incorporated as a tower of general covariant operators localized on the particle's worldline. We can then apply the machinery of worldline EFT to compute scattering amplitudes and observables from the path integral

\begin{equation}
    Z=\int \mathcal{D}[h]\mathcal{D}[A]\mathcal{D}[W]e^{iS_{\rm bulk}+iS+iS_{\rm gf}}.
\label{eq:pathintegral}
\end{equation}
Here, $S_{\rm bulk}, S, S_{\rm gf}$ are the bulk action describing the graviton and photon field self-interactions, the worldline action describing the spinning charged particle as well as its interaction with the boson fields, and the gauge-fixing action. By expanding the space-time metric around a flat background $g_{\mu\nu}=\eta_{\mu\nu}+h_{\mu\nu}$, we treat the metric perturbation $h_{\mu\nu}$ as a propagating spin-2 quantum field ``graviton.'' $A_\mu$ is the photon field, and $W=\{z^\mu,\pi^\mu,s^{\mu\nu},\lambda^{\mu\nu}\}$ denotes the full set of worldline fluctuations due to the interaction with the metric perturbation and external electromagnetic field, which we will describe in full detail when we define and gauge fix the worldline action $S$. We define the action terms explicitly as follows.

The bulk gravitational and electromagnetic fields are described by the Einstein-Hilbert and Maxwell action:
\begin{align}
S_{\rm bulk} &= \int d^4x\left[-\frac{1}{2\kappa}\sqrt{-g}\,R - \frac14 \sqrt{-g}\,F^2\right],
\label{eq:bulk}
\end{align}
where $\kappa=8\pi G$, $R$ is the Ricci scalar, and $F^2=F^{\mu\nu}F_{\mu\nu}$ is the covariantly contracted Maxwell tensors. We fix the gauge of the boson fields by adding a gauge-fixing term to the bulk action:
\begin{align}
    S_{\rm gf}&=\int d^4x(\partial_\nu h^{\mu\nu}-\frac{1}{2}\partial_\mu h)^2-\frac{1}{2}  \, \left(\partial^\mu A_\mu \right)^2,
\end{align}
where the first term enforces the de Donder gauge for the graviton while the second term chooses the Feynman gauge for the photon. The propagator of the graviton field is derived by inverting the kernel of the $\mathcal{O}(h^2)$ term in the action, and is given by

\begin{subequations}
\label{eq:hA propagators}
\begin{align}
D^h_{\mu\nu,\rho\sigma}(k)
= \frac{i}{k^2 + i\epsilon}\;
\mathcal{P}^h_{\mu\nu,\rho\sigma},
\qquad\\
\mathcal{P}^h_{\mu\nu,\rho\sigma}
= \tfrac12\!\left(
\eta_{\mu\rho}\eta_{\nu\sigma}
+\eta_{\mu\sigma}\eta_{\nu\rho}
-\eta_{\mu\nu}\eta_{\rho\sigma}
\right).
\end{align}
\end{subequations}
The Feynman gauge photon propagator reads: 
\begin{equation}
    \mathcal{D}^A_{\mu\nu}=\frac{-i\eta_{\mu\nu}}{k^2+i\epsilon}.
\end{equation}
The higher-order terms introduce an infinite number of graviton self-interactions and graviton-photon mixing interactions. In particular, there is an $hAA$ coupling that contributes in our tree-level calculation. Higher order terms are necessary for extending to loop calculations, and may contribute in other processes. 

To describe the spin coupling naturally defined in the local Lorentz frame, we introduce a tetrad $e^\mu_a$ with the Greek indices for the general spacetime manifold and Latin indices for the local Lorentz frame. The choice of the local frame can be transformed under a $SO(3)$ little group, which corresponds to a spatial rotation at a fixed time slice. This rotation can be formalized by defining a local Lorentz matrix $\Lambda^a_I(\tau)$ on the worldline where the capital Latin indices are for the co-rotating, body-fixed local frame. The angular velocity 
can thus be written in terms of the Lorentz matrix as 
\begin{equation}
    \Omega^{ab}=\Lambda^{aI}\frac{D\Lambda^b_I}{d\tau}.
\end{equation}

Since the unconstrained spin tensor $S_{ab}$ and Lorentz matrix $\Lambda_a{}^{\mu}$ each contain six independent components, one must impose a spin-supplementary condition (SSC) to eliminate the unphysical degrees of freedom associated with the choice of body-fixed frame. In the covariant SSC, this is done by imposing
\begin{equation}
    S_{ab}\hat{p}^b = 0,
    \qquad
    \Lambda_{a0} = \hat{p}_a,
\end{equation}
where
\begin{equation}
    \hat{p}^\mu \equiv \frac{p^\mu}{\sqrt{p^2}}.
\end{equation}
These constraints remove the three redundant boost-like components, leaving three physical rotational degrees of freedom in the spin sector. Thus both $S_{ab}$ and $\Lambda_a{}^{\mu}$ are reduced to their physical three-component content. 

Instead of applying the SSC directly, we follow~\cite{Levi_2015, benshahar2023scatteringspinningcompactobjects} by first applying a single constraint at the cost of introducing a gauge symmetry, and later gauge-fix with the Lagrange multiplier. The worldline action in a first-order form is then defined as:
\begin{align}
\label{eq:master_action}
    S&=-\int d\tau \Big[p_\mu \dot{x}^\mu+\frac{1}{2}S_{ab}\Omega^{ab}+eA_\mu\dot{x}^\mu+\frac{1}{p^2}\frac{Dp_\mu}{d\tau}S^{\mu\nu}p_\nu \nonumber\\
    &\quad
    -\frac{l}{2}(p^2-m^2-\mathcal{L}_{\text{n.m.}})-l_aS^{ab}(\hat{p}_b+\Lambda_{0b})\Big]
\end{align}
where $l$, $l_a$ are the Lagrange multipliers, and the non-minimal operators have been incorporated under the mass-shell condition. $l$ encodes the reparameterization invariance, while $l_a$ controls the SSC gauge fixing. The constraints enforced by the Lagrange multipliers in Eq. (\ref{eq:master_action}) must be first-class to be of physical significance, which we have verified with the electromagnetic interactions included in the action, in Appendix~\ref{app:eomsandconstraint}. In particular, $l_a$ produces the gauge symmetry defined by the following gauge transformations~\cite{benshahar2023scatteringspinningcompactobjects,steinhoff2015spingaugesymmetryaction}:
\begin{subequations}
\label{eq:gauge transformations}
\begin{align}
\delta S^{\mu\nu} &= 2 \hat{p}^{[\mu} S^{\nu]\alpha} \, \epsilon_\alpha , \\
\delta \Lambda^I_{\mu} &= 2 \epsilon_{[\mu} \hat{p}_{\nu]} \Lambda^{I\nu} 
  + 2 \epsilon_{[\mu} \Lambda_{\nu]0} \Lambda^{I\nu} ,\\
\delta l^\mu&=-\frac{D\epsilon^\mu}{d\tau}+\cdots \, .
\end{align}
\end{subequations}
We may conveniently fix the spin gauge and reparameterization invariance by choosing 
\begin{subequations}
\begin{align}
\label{eq:gauge-fixing}
    l_\mu&=\frac{1}{p}\frac{Dp_\mu}{d\tau},
    \qquad\\
    &l=\frac{1}{m}.
\end{align}
\end{subequations}
Here we have adopted the proper time parameterization. Under our SSC choice, $\Lambda_{0\mu}=\hat{p}_\mu$ for the asymptotic states, which corresponds to the covariant SSC and replaces the single first-class constraint with two second-class constraints. The gauge-fixed worldline action thus reads
\begin{align}
\label{eq:gauge-fixed action}
    S&=-\int d\tau \Big[p_\mu \dot{x}^\mu+\frac{1}{2}S_{ab}\Omega^{ab}+eA_\mu\dot{x}^\mu \nonumber\\
    &\quad
    -\frac{1}{2m}(p^2-m^2)+\mathcal{L}_{\text{n.m.}}-\frac{1}{p}\frac{Dp_\mu}{d\tau}S^{\mu\nu}\Lambda_{\nu0}\Big],
\end{align}
where we have absorbed the constant $1/2m$ factor from the reparameterization gauge fixing Lagrange multiplier into the Wilson coefficients for the non-minimal operators. 

The more complicated internal structure or background geometry of the compact object is described by non-minimal operators. These curvature and EM-field-dependent operators should respect both general gauge and $U(1)$ gauge invariance. Non-minimal spin-dependent operators should also be invariant under gauge transformations in Eq.~(\ref{eq:gauge transformations}) in order to consistently propagate the three degrees of freedom in the spin sector~\cite{benshahar2023scatteringspinningcompactobjects}. One clean way to achieve the spin-gauge invariance is to build any spin-dependent operators only in terms of the spin vector projected by $\hat{p}^\mu$, $S^\mu=\frac{1}{2}\epsilon^{\mu\nu\rho\sigma}\hat{p}_\nu S_{\rho\sigma}$. Alternatively, one may define the operator in a form where spin is encoded by the projected spin tensor $\hat{S}^{\mu\nu}=S^{\rho\sigma}(\delta^\mu_\rho - \hat{p}^\mu \hat{p}_\rho) (\delta^\nu_\sigma - \hat{p}^\nu \hat{p}_\sigma)$. To build the gravitational operators, we define the electric and magnetic parity components of the Riemann tensor as~\cite{Goldberger:2006bd}:
\begin{align}
\label{eq:Riemann projection}
E_{\mu\nu} &= R_{\mu\rho\nu\sigma} \hat{p}^{\rho} \hat{p}^{\sigma}, \\
B_{\mu\nu} &= \tfrac{1}{2} R_{\alpha\beta\rho\mu} \, 
  \epsilon^{\alpha\beta}{}_{\gamma\nu} \, \hat{p}^{\rho} \hat{p}^{\gamma}.
\end{align}
For purely gravitational couplings, we thus have non-minimal linear-in-Riemann couplings
\begin{align}
\label{eq:gravitational couplings}
\mathcal{O}(R S^{2n}) &\sim D_{\mu_{2n}} \cdots D_{\mu_3} 
   E_{\mu_1 \mu_2} S^{\mu_1} \cdots S^{\mu_{2n}}, \\
\mathcal{O}(R S^{2n+1}) &\sim D_{\mu_{2n+1}} \cdots D_{\mu_3} 
   B_{\mu_1 \mu_2} S^{\mu_1} \cdots S^{\mu_{2n+1}}.
\end{align}
For the EM couplings, we also define analogously the electric and magnetic part of the field tensor $F_{\mu\nu}$ projected with the $\hat{p}^\mu$ vector:
\begin{align}
\label{eq:F projection}
E_{\mu} &= F_{\mu\nu}
\hat{p}^{\nu}, \\
B_{\mu} &= \tfrac{1}{2} \, 
  \epsilon_{\mu\nu}{}^{\alpha\beta} F_{\alpha\beta}\, \hat{p}^\nu.
\end{align}
We can then organize the non-minimal EM couplings as
\begin{align}
\label{eq:EM couplings}
\mathcal{O}(F S^{2n}) &\sim D_{\mu_{2n}} \cdots D_{\mu_2} 
   E_{\mu_1} S^{\mu_1} \cdots S^{\mu_{2n}}, \\
\mathcal{O}(F S^{2n+1}) &\sim D_{\mu_{2n+1}} \cdots D_{\mu_2} 
   B_{\mu_1} S^{\mu_1} \cdots S^{\mu_{2n+1}}.
\end{align}

The operators are naturally structured as an expansion in $\epsilon=r_s/\lambda\sim\omega m$, where $\omega$ is the typical energy of the external photon/graviton. We work in the long wavelength regime, where $\epsilon \ll 1$. This unique power counting scheme allows us to organize the calculations up to a certain order in $\epsilon$. For clarity, for loop contributions beyond the tree level, the quantum corrections will be counted by powers of $l_{\rm pl}/\lambda$, where $l_{\rm pl}$ is the Planck length scale. For a macroscopic black hole, $\frac{l_{\rm pl}/\lambda}{\epsilon}\sim m_{\rm pl}/m\ll 1$, hence the quantum corrections are highly suppressed.

In this work, we compute the graviton photoproduction scattering amplitude up to $\mathcal{O}(\epsilon^2)$ at tree level, which produces the first non-trivial self-consistent differential cross section in a spin expansion through $\mathcal{O}(S^2)$. Through this order, mixed local operators schematically written as $RF$ do not contribute. Here $R$ stands for the Riemann tensor and $F$ stands for the Maxwell tensor, with contractions and possible dependence on the worldline variables such as $v^\mu$, $S^\mu$ implicit. The Riemann tensor contains two derivatives that translate to $\mathcal{O}(\epsilon^2)$, while the Maxwell contains one derivative contributing at $\mathcal{O}(\epsilon)$. Together with the natural length scale of our worldline EFT $r_s$, such operators first contribute at $\mathcal{O}(\epsilon^3)$, even without additional covariant derivatives. Given the intended order of our calculation, such EM-gravity mixed operators are at least suppressed by a factor of $\epsilon$. For the graviton photoproduction scattering amplitude, this suppression at long wavelengths suggests that the differential cross section up to out working order is universally fixed by the multipole moments. Since the mixing operators do not contribute to the observables of interest in our analysis, they will be safely dropped in the construction of $\mathcal{L}_{n.m.}$. 

Although we focus on the KN black hole in this study, we emphasize that the framework employed in this work applies more generally to EM-gravity interactions with spinning charged particles. Beyond the order we consider, the worldline EFT needs to be matched to the specific object under analysis with a correctly chosen operator basis. For example, for a generic neutron star, the local quadratic operators must in general be included, and their Wilson coefficients are determined by matching to the ultraviolet physics describing the internal structure specific to the object under consideration. 

The non-conservative effects do not contribute at the order of interest here, but can be incorporated systematically within the same worldline framework in future extensions. Schematically, such effects may be captured through the non-local quadratic response operators of the form
\begin{equation}
    \chi_{RF}(\omega)\,RF, 
\end{equation}
where $R$ and $F$ denote the curvature and Maxwell tensors, respectively, with index contractions left implicit. ``$RF$'' together contributes $\mathcal{O}(\epsilon^3)$ as we have shown. Here $\chi_{RF}(\omega)$ is a frequency-domain retarded Green's function encoding the dynamical response of the compact object. Such coefficients naturally emerge in an in-in formulation of the worldline EFT, where they represent retarded response in an external field~\cite{Goldberger:2020fot, Galley:2009px}. The non-conservative effects are in turn described by the imaginary parts. As a result, non-conservative effects are further parametrically suppressed in the long-wavelength regime relative to the conservative contributions, and therefore do not enter the conversion amplitude through at least $\mathcal{O}(\epsilon^3)$. Consequently, the amplitude computed to the order $\mathcal{O}(\epsilon^2)$ remains conservative and elastic.

We proceed to identify explicitly the non-minimal operators that are relevant to the intended order. On the gravity side, the leading order contribution is $E_{\mu\nu}S^\mu S^\nu$. Other purely gravitational operators will be suppressed by powers of $\epsilon$. For EM operators, the leading order operator relevant for a KN black hole power counted in either $\epsilon$ or $S$ is $B_\mu S^\mu$, followed by $D_\mu E_\nu S^\mu S^\nu$ suppressed by an additional power of $\epsilon$. In general, there could also be an electric dipole operator, which we denote as $E_\mu d_{\text{EM}}^\mu$, where $d_{\text{EM}}$ is the electric dipole. Collecting the terms, up to $\mathcal{O}(\epsilon^2)$ (and correspondingly $\mathcal{O}(S^2)$), the non-minimal couplings read

\begin{align}
\label{eq:non-minimal L}
    \mathcal{L}_{\text{n.m.}}&=lc_1E_\mu d_{\text{EM}}^\mu+lc_2B_\mu S^\mu+\frac{lc_3}{2m}D_\mu E_\nu S^\mu S^\nu
    \nonumber\\
    &\quad
    +\frac{lc_4}{2}E_{\mu\nu}S^\mu S^\nu.
\end{align}

The operators controlled by $c_2$, $c_3$, and $c_4$ encode the magnetic dipole moment, electric quadrupole moment, and mass quadrupole moment respectively. The expansion $\epsilon$ coincides with an expansion in the spin vector $S^\mu$, which is specific to a KN black hole, where the EM and mass multipoles are induced by the spin in a known form. For instance, the magnetic dipole is parallel to the spin, and the strength and direction of the magnetic moment are characterized by the gyromagnetic ratio $g_{\text{EM}}$ which happens to be 2. For a more general compact object, the angular momentum is not necessarily aligned with the magnetic dipole, hence the expansion in terms of spin will no longer apply automatically. Instead, the EFT should be power counted strictly in terms of $\epsilon$.

To account for the worldline fluctuations by interacting with the external fields, we expand the worldline degrees of freedom around the solutions to flat spacetime vacuum equations of motion for the worldline:
\begin{subequations}
\begin{align}
    &x^\mu\to b^\mu+v^\mu\tau+z^\mu,\\
    &p_\mu\to mv_\mu+\pi_\mu,\\
    &S_{ab}\to S_{ab}+s_{ab},\\
    &\Lambda^a_I\to \Lambda^a_I+\lambda^{ab}\Lambda_{Ib}+\frac{1}{2}\lambda^{ab}\lambda_{bc}\Lambda^c_I+...~.
\end{align}
\end{subequations}
Here $x^\mu$ is the worldline coordinate, $p_\mu$ is the momentum, $S_{ab}$ is the spin tensor in the local Lorentz frame, and $\Lambda^a_I$ is the Lorentz matrix. We normalize the constant $v_\mu$ by $v^2=1$, and take the asymptotic states to obey the covariant SSC: $\Lambda^\mu_0=v^\mu$ and $S_{\mu\nu}v^\nu=0$. $S_{ab}$ and $\Lambda^a_I$ of the asymptotic states are constants in accordance to the internal spin of the particle.

\section{Matching to Kerr-Newman Black Hole}
\label{sec:matching}
Our analysis in the previous section shows that the conservative graviton-photon conversion amplitude is completely determined by the Kerr-Newman multipole structure up to $\mathcal{O}(S^2)$ in the long wavelength regime. Non-conservative effects and higher-derivative mixed operators involving both curvature and EM  field strength introduce additional powers of frequency, and therefore are suppressed. As a result, no matching to Teukolsky scattering solutions is required at this order. To determine the Wilson coefficients, we match the potential computed from our worldline EFT operators to the Kerr-Newman metric and Maxwell field at spatial infinity, consistent with our long-wavelength assumption. 
\subsection{EM Multipole}
The EM multipole structure of a Kerr-Newman black hole is well known. In Boyer-Lindquist coordinates, the EM potential is 
\begin{align}
A_\mu=\big(-\frac{er}{\rho^2},0,0,\frac{e\ ra\sin^2(\theta)}{\rho^2}\big),
\end{align}
where $a=S/m$, $\rho^2=r^2+a^2 \cos^2(\theta)$. To compare with the far zone fields computed from the worldline EFT, we need to transform into the asymptotically Cartesian and mass-centered ($\text{ACMC}$) coordinates~\cite{RevModPhys.52.299}. The transformation can be achieved by 
\begin{align}
    r'\sin\theta'=\sqrt{r^2+a^2}\sin\theta,\ r'\cos\theta'=r\cos\theta,
\end{align}
where the primed coordinates are in the $\text{ACMC}-\infty$ coordinates. The Maxwell potential at the spatial infinity in $\text{ACMC}-\infty$ coordinates becomes~\cite{Ma_2025}
\begin{align}
\label{eq:EM multipole}
\notag
A_t &=  -\sum_{\ell=0}^{\infty} \Big(\frac{1}{r^{\ell+1}} \, Q_\ell \, P_\ell(\cos\theta)+\sum_{\ell'<\ell}c^{(t)}_{\ell\ell'}P_{\ell'}\Big),\\
\qquad
A_\varphi &=\sum_{\ell=0}^{\infty} \frac{\cos\theta}{r^{2\ell}} \,\Big(  M_{2\ell} \, P_{2\ell}(\cos\theta)+\sum_{\ell'<\ell}c^{(\varphi)}_{2\ell, 2\ell'}P_{2\ell'}\Big)\\
\notag
&-\sum_{\ell=0}^{\infty}\frac{1-\cos^2\theta}{r^{2\ell+1}} \,\frac{1}{2\ell+1} \, \Big(M_{2\ell+1} \, P'_{2\ell+1}(\cos\theta) \\
&+\sum_{\ell'<\ell}c^{(\varphi)}_{2\ell+1,2\ell'+1}P'_{2\ell'+1}\Big),
\end{align}
where $P_\ell$ is the Legendre polynomial and $P'_\ell$ is the x derivative of $P_\ell(x)$, $c$'s are gauge-dependent numerical coefficients. The gauge terms are non-physical and do not interfere with the physical multipole moments. We take advantage of this clear separation from the referenced expansion in the following discussion. Here $Q_\ell$ denotes the electric multipole and $M_\ell$ denotes the magnetic multipole, given by 
\begin{align}
    Q_{2n}&=e(-a^2)^n, \, Q_{2n+1}=0\\
    \notag
    M_{2n+1}&=-ea(-a^2)^n, \, M_{2n}=0,
\end{align}
where $a=S/m$. 

It follows that $c_1=0$, namely there is no electric dipole operator. To match the Pauli operator
\begin{align}
S\supset-\int d\tau\, l c_2 B_\mu S^\mu,
\end{align}
we evaluate the fields generated by the EFT operators in the black hole rest frame, consistent with the asymptotically inertial, mass-centered ACMC coordinates used to define the multipole moments, so that
\begin{align}
    J^j&=\frac{\delta}{\delta A_i}\Big(\frac{c_2}{m}S_kB^k\delta^{(3)}(\vec{x})\Big)=-\partial_i\big(\frac{c_2}{m}\epsilon^{ijk}S_k\delta^{(3)}(\vec{x})\big)\notag\\
    &=\big(\vec{\nabla}\times\frac{c_2}{m}\vec{S}\delta^{(3)}(\vec{x})\big)^j.
\end{align}
In our choice of Lorenz (Feynman) gauge and by virtue of the stationarity of the solution, by solving the Poisson equation $\nabla^2\vec{A}=-\vec{J}$, we get
\begin{align}
    A_i(\vec{r})=\frac{c_2}{m}\epsilon_{ijk}S_j\frac{x_k}{r^3}.
\end{align}
Hence
\begin{align}
    A_\varphi=\frac{c_2}{m}\frac{S\sin^2\theta}{r}.
\end{align}
To extract the magnetic dipole induced $A_{\varphi}$, we observe that this corresponds to the $\ell=0$ terms in the expansion in Eq.~(\ref{eq:EM multipole}). The potential corrections induced by the gauge choices identically vanish at this order, hence do not confuse with the matching of the magnetic dipole coefficient. Comparing with Eq. (\ref{eq:EM multipole}), we obtain $c_2=e$. 

To match the electric quadrupole coupling, we observe that
\begin{align}
    \rho(\vec{x})&=-\frac{\delta }{\delta \phi}\Big(-\frac{c_3}{2m^2}S^iS^j\partial_iE_j\delta^{(3)}(\vec{x})\Big)\notag\\
    &=-\frac{c_3}{2m^2}\frac{\delta}{\delta\phi}\Big(S^iS^j\partial_i\partial_j\phi\delta^{(3)}(\vec{x})\Big)\notag\\
    &=-\frac{c_3}{2m^2}S^i S^j\partial_i \partial_j \, \delta^{(3)}(\vec{x}),
\end{align}
and by solving the Poisson equation $\nabla^2\phi=-4\pi\rho$, we have
\begin{align}
    A_t&=-\phi=(-1)^2\frac{c_3}{2m^2}S^i S^j\partial_i\partial_j(\frac{1}{r})\\
    \notag&=\frac{c_3}{2m^2}\frac{S^2}{r^3}(3\cos^2\theta-1).
\end{align}
This should correspond to the $\ell=2$ terms in the expansion in Eq.~(\ref{eq:EM multipole}). At this order, the gauge dependent correction is a linear combination of $P_0(\cos\theta)=1$ and $P_1(\cos\theta)=\cos\theta$, which does not spoil the uniqueness of the potential's spatial dependence induced by the physical electric quadrupole moment. Matching the EFT derived potential with Eq. (\ref{eq:EM multipole}), we obtain $c_3=e$. 

\subsection{Mass Multipole}
In ACMC-2 coordinates, the Kerr-Newman metric takes the following form by coordinate transformation from the Boyer-Lindquist coordinates
\begin{align}
\label{eq:ACMC transform}
r = r' + \frac{a^2 \cos^2 \theta'}{2r'}, \quad 
&\theta = \theta' - \frac{a^2 \cos \theta' \sin \theta'}{2r'^2}, \quad 
\phi = \phi', \\\notag \quad 
&t = t',
\end{align}
and reads~\cite{RevModPhys.52.299} 
\begin{align}
\label{eq:KN metric_00}
g_{0'0'} &= 1 - \frac{2m}{r'} - \frac{e^2}{r'^2}
           + \frac{3ma^2 \cos^2 \theta'}{r'^3} + \mathcal{O}\!\left(\frac{1}{r'^5}\right)\\
           \notag&= 1 - \frac{2m}{r'} - \frac{e^2}{r'^2}
           + \frac{2ma^2}{r'^3}\big[P_2(\cos\theta')+\frac{1}{2}\big]+ \mathcal{O}\!\left(\frac{1}{r'^5}\right).
\end{align}
Starting from the worldline action
\begin{align}
S\supset-\int d\tau \frac{c_4}{2m}E_{\mu\nu}S^\mu S^\nu,
\end{align}
we have in the rest frame of the particle
\begin{align}
T_{00}=\frac{c_4}{2m}\partial_i\partial_j\delta^{(3)}(\vec{x})S^iS^j.
\end{align}
The far zone stationary solution of the gravitational field is thus given by solving the Poisson equation $\nabla^2h_{00}=-8\pi G\, T_{00}$, which yields
\begin{align}
    h_{00}(\vec{x})&=\frac{G\,c_4}{m}S^iS^j\partial_i \partial_j\big(\frac{1}{r}\big)\\
    \notag&=\frac{G\,c_4S^2}{mr^3}\big(3\cos^2(\theta)-1\big)\\
    \notag&=\frac{2c_4G\, S^2}{mr^3}P_2(\cos\theta).
\end{align}
For any different $\text{ACMC-N}$ coordinates and $\ell\leq N+1$, the far zone potentials induced by the physical mass multipole $\ell$-moments have exactly the same form~\cite{RevModPhys.52.299}. Meanwhile, the de Donder gauge corresponds to a $\text{ACMC}-\infty$ coordinate system, hence the multipole moments calculated from our EFT action should agree with the multipole moments derived in Eq.~(\ref{eq:KN metric_00}). Comparing this with Eq. (\ref{eq:KN metric_00}), where $G=1$, $a=S/m$, we get $c_4=1$. This result is the same as the case for a Kerr black hole without EM interactions, and does not depend on the electric charge. It is also consistent with the Wilson coefficient derived by matching with the Teukolsky solutions directly~\cite{bautista2023scatteringblackholebackgrounds1, bautista2023scatteringblackholebackgrounds2, benshahar2023scatteringspinningcompactobjects}.

\section{Feynman rules}
\label{sec:feynmanrules}
Having established the operator basis, we now turn to the derivation of the Feynman rules used in the amplitude calculation. In this section, we identify the relevant propagators and interaction vertices and show how they follow from the action.

\subsection{Propagators}
In addition to the boson field propagators, we need the propagators of the worldline fluctuation fields $W$. This is equivalent to keeping track of the full equations of motion with external fields interacting with the worldline. The quadratic in $W$ and zeroth order in boson fields terms are:
\begin{align}
S_{\text{kin}} = - \int d\tau \Big[
    & \, \dot{z}^\mu \pi_\mu 
    - \frac{1}{2m}\pi^2
    + \frac{1}{2} S^{\mu\nu}\lambda_{\mu\rho}\dot{\lambda}^{\nu\rho}
    - \frac{1}{2} \dot{\lambda}^{\mu\nu} s_{\mu\nu} \nonumber \\
    & \; - \frac{1}{m} S^{\mu\nu}\pi_\mu \lambda_{\nu\rho} v^\rho
    - \frac{1}{m} \dot{\pi}^\mu s_{\mu\nu} v^\nu
\Big] \, .
\end{align}
By inverting the kernel of the quadratic terms and working in the energy/momentum space ($z^\mu(\tau)=\int d\omega/2\pi ~e^{i\omega t}z^\mu(\omega)$, etc.), we compute the two-point functions of the worldline fluctuation fields as:
\begin{subequations}
\begin{align}
\langle z^\mu(-\omega) z^\nu(\omega) \rangle 
&= - i \frac{1}{m\omega^2} \eta^{\mu\nu} 
   - \frac{1}{m^2 \omega} S^{\mu\nu} , \label{eq:a}\\[6pt]
\langle \pi^\mu(-\omega) z^\nu(\omega) \rangle 
&= - \frac{1}{\omega} \eta^{\mu\nu} , \label{eq:b}\\[6pt]
\langle s_{\mu\nu}(-\omega) s_{\rho\sigma}(\omega) \rangle 
&= - \frac{2}{\omega} 
   \big( \eta_{\nu[\sigma} S_{\rho]\mu} 
       - \eta_{\mu[\sigma} S_{\rho]\nu} \big) , \label{eq:d}\\[6pt]
\langle s_{\mu\nu}(-\omega) \lambda_{\rho\sigma}(\omega) \rangle 
&= \frac{2}{\omega} \, \eta_{\mu[\rho} \eta_{\sigma]\nu} , \label{eq:e}\\[6pt]
\langle \lambda^{\mu\nu}(-\omega) z_\rho(\omega) \rangle 
&= - \frac{2}{m\omega} \, v^{[\mu} \delta^{\nu]}_{\rho} . \label{eq:f}
\end{align}
\label{eq: W propagators}
\end{subequations}
To get the retarded(advanced) propagators, one may use the usual $\frac{1}{\omega}\to \frac{1}{\omega \pm i0}$ replacement. In the current calculation we use the average of the far past and far future background fields~\cite{Mogull_2021, Galley:2009px}. The propagators for the graviton and photon fields are given in Eq.~(\ref{eq:hA propagators}).

\subsection{Vertices}
The boson field interactions can be derived from the bulk action. There are two types of interaction vertices that can be read off $S_{\rm bulk}$ defined in Eq.~(\ref{eq:bulk}). First, there is an arbitrary number of gravitons connected to a graviton self-interaction vertex, which come from the Einstein-Hilbert term after expanding the Ricci scalar $R$ and metric determinant $\sqrt{-g}$ in terms of $h$. The photon-graviton mixing vertices come from the second term, with two photon legs and an arbitrary number of graviton legs starting from 1. For the tree level amplitude considered here, we need only the $hAA$ vertex. For scattering processes with different in and out states or the loop calculations, the higher order vertices are generally needed as well, but the derivation follows the same procedure. We work in the momentum space, where the Fourier transform is defined as 
\begin{subequations}
\begin{align}
    &h_{\mu\nu}(x)=\int \frac{d^4k}{(2\pi)^4}e^{ik\cdot(b+v\tau+z)}h_{\mu\nu}(k),\\
    &A_\mu(x)=\int \frac{d^4k}{(2\pi)^4}e^{ik\cdot(b+v\tau+z)}A_\mu(k).
\end{align}
\end{subequations}
Substituting this into the action in Eq.~(\ref{eq:bulk}), we derive the $hAA$ vertex rule to be 
\begin{align}
\label{eq:hAA}
&V_{hAA}^{\mu\nu;\alpha\beta}(k_1,k_2) 
= \frac{i}{2} \Big[
   (k_1 \!\cdot\! k_2)\,(\eta^{\mu(\alpha}\eta^{\beta)\nu}-\eta^{\mu\nu}\eta^{\alpha\beta}) \nonumber \\
&\quad
   + \eta^{\mu\nu} k_1^{(\alpha} k_2^{\beta)}
   + \eta^{\alpha\beta} k_1^{(\mu} k_2^{\nu)}
   - \eta^{\mu(\alpha} k_2^{\beta)} k_1^{\nu} 
   \nonumber \\
&\quad
   - \eta^{\nu(\alpha} k_2^{\beta)} k_1^{\mu}
   - \eta^{\mu(\alpha} k_1^{\beta)} k_2^{\nu}
   - \eta^{\nu(\alpha} k_1^{\beta)} k_2^{\mu}
   \Big] ,
\end{align}
where we take all the momenta to be incoming, and have the graviton leg $h_{\mu\nu}$ as well as photon legs $A_\alpha(k_1)$, $A_\beta(k_2)$.

To compute the worldline coupling vertices, we expand the worldline action in Eq.~(\ref{eq:gauge-fixed action}) around the flat spacetime vacuum background and an unperturbed worldline. We consider the graviton vertices first, and expand to $\mathcal{O}(S^2)$. For completeness, we compute the single graviton emission vertex first. Although this rule is not used in the current calculation, it becomes relevant in loop computations and other processes such as the photon-to-photon scattering sourced by the gravitational background of the black hole. 

\begin{align}
\begin{tikzpicture}[baseline=(current bounding box.center)]
  \draw[dotted] (-0.75,0) -- (0.75,0);
  \draw[graviton] (0,0) -- (0,-1);
\end{tikzpicture}
&= -\frac{i}{2} m (v \cdot h \cdot v)
   - \frac{1}{2} (v \cdot h \cdot S \cdot k)\notag\\
   &-i\mathcal{L}_{n.m.}\,\big|_h .
\end{align}

The external graviton can also source a worldline fluctuation. Alternatively, the propagated worldline fluctuation can lead to the emission of a graviton. Such processes are captured by the vertex rule:
\begin{align}
&\begin{tikzpicture}[baseline=(current bounding box.center)]
  % dotted incoming line
  \draw[dotted,line width=0.4pt] (-0.75,0) -- (0,0);
  % short solid ledge
  \draw[line width=0.9pt] (0,0) -- (0.75,0);
  % downward wavy line
  \draw[graviton]
       (0,0) -- (0,-1);
\end{tikzpicture}
=\frac{m}{2}\,(v \cdot h \cdot v)(k \cdot z)
- i\,(\pi \cdot h \cdot v) \notag \\[6pt]
&\quad - \frac{i}{2}\,(v \cdot h \cdot S \cdot k)(z \cdot k) 
- \frac{1}{2}\,(v \cdot h \cdot s \cdot k) \notag \\[6pt]
&\quad + \frac{i}{2}\,(z \cdot h \cdot S \cdot k)\,(k\cdot v) + \frac{1}{2}\,(v \cdot h \cdot v)(v \cdot \lambda \cdot S \cdot k) \notag \\[6pt]
&\quad - \frac{1}{2}\,(v \cdot h \cdot v)(v \cdot s \cdot k)-i\mathcal{L}_{n.m.}\,\big|_{h, W}
\end{align}
where the worldline fluctuations are expanded to linear order. 

The photon vertices have the same topology, and the single photon insertion vertex up to $\mathcal{O}(\epsilon^2)$ is
\begin{align}
\begin{tikzpicture}[baseline=(current bounding box.center)]
  \draw[dotted] (-0.75,0) -- (0.75,0);
  \draw[decorate,decoration={snake,amplitude=1.2pt,segment length=6pt},line width=0.8pt] (0,0) -- (0,-1);
\end{tikzpicture}
= &-ie(v\cdot A)-\frac{c_2}{m}(k\cdot S\cdot A)-i\mathcal{L}_{n.m.}\,\big|_A.
\end{align}

The photon vertex that sources a worldline fluctuation is
\begin{align}
\begin{tikzpicture}[baseline=(current bounding box.center)]
  % dotted incoming line
  \draw[dotted,line width=0.4pt] (-0.75,0) -- (0,0);
  % short solid ledge
  \draw[line width=0.9pt] (0,0) -- (0.75,0);
  % downward wavy line
  \draw[decorate,decoration={snake,amplitude=1.2pt,segment length=6pt},line width=0.8pt]
       (0,0) -- (0,-1);
\end{tikzpicture}
&=e(v\cdot A)(k\cdot z)+e(z\cdot A)\omega-i\mathcal{L}_{n.m.}\,\big|_{A, W}
\end{align}
Finally, covariantizing the EM non-minimal operators generates the seagull terms connecting to both a graviton and a photon. More specifically, the graviton leg comes from transforming the spin tensors $S_{ab}$ from the local Lorentz frame to the general spacetime coordinates using the tetrad, and explicitly expanding the contractions with $g_{\mu\nu}=\eta_{\mu\nu}+h_{\mu\nu}$. The covariant derivatives also contribute to the seagull terms by virtue of the Christoffel symbols. The complete rule is attached in Appendix~\ref{app:feynman rules}, and we include the subset rules proportional to $c_2$ here:
\begin{align}
\begin{tikzpicture}[baseline=(current bounding box.center)]
  % dotted incoming line
  \draw[dotted,line width=0.4pt] (-0.75,0) -- (0.75,0);
  % downward wavy line
  \draw[graviton]
       (0,0) -- (-0.75,-1);
  % downward wavy line
  \draw[decorate,decoration={snake,amplitude=1.2pt,segment length=6pt},line width=0.8pt]
       (0,0) -- (0.75,-1);
\end{tikzpicture}
&=\frac{c_2}{2m}\Big[-(k_A\cdot h\cdot S\cdot A)+(A\cdot h\cdot S\cdot k_A)\notag\\
&-(v\cdot h\cdot S\cdot k_A)(A\cdot v)+(v\cdot h\cdot S\cdot A)(k_A\cdot v)\Big]\notag\\
&+\dots \notag\\[6pt]
\end{align}

In all the Feynman rules, we use incoming momenta/energy for all particles and worldline fluctuations. We use double wavy lines for the graviton, single wavy lines for the photon, dotted lines for the worldline and solid lines for the worldline fluctuations. At each vertex, an energy conserving delta function $2\pi\delta(k_1\cdot v+k_2\cdot v)$ or $2\pi\delta(k\cdot v+\omega)$ is implicitly assumed. We also suppress the coupling $\kappa$ in the graviton rules. In the seagull diagram for the contact terms derived from individual operators, $k_A$, $k_h$ denote the four-momenta of the photon and graviton respectively. Furthermore, we use ``$\cdot$'' as an abbreviated notation for index contractions. For example, $S^{\mu\nu}k_\mu A_\nu$ is written as $k\cdot S\cdot A$. The full vertex rules including the non-minimal contributions can be found in Appendix~\ref{app:feynman rules}. In computing the scattering amplitude, the vertices that connect to a worldline fluctuation are contracted with the non-trivial worldline fluctuation propagators defined in Eq.~(\ref{eq: W propagators}).

\twocolumngrid

\section{Results}
\label{sec:results}
In this section, we present the tree-level calculation of the graviton photoproduction amplitude by a KN black hole, $g\to\gamma$, in the low-energy regime through $\mathcal{O}(\epsilon^2)$. The result is then subjected to nontrivial consistency checks. In particular, we verify that the amplitude satisfies both the gauge invariance associated with the external bosonic fields and the spin-gauge invariance associated with the SSC, and reduces to the graviton photoproduction amplitude by a RN black hole in the spinless limit. We then compute the corresponding differential cross section and analyze the spin-dependent corrections through $\mathcal{O}(S^2)$. This provides the first explicit characterization of how the KN black hole background modifies graviton--photon scattering in a long-wavelength expansion. 

Although formulated as a one-graviton to one-photon scattering process, the tree-level result is classical in content. The massive source is treated as a classical worldline, and the tree diagrams simply encode the solution of its classical equations of motion and the induced response to the radiative fields. Quantum corrections can be incorporated systematically within the same framework by computing loop diagrams of the graviton and photon fields.

\subsection{Amplitudes}
Our main result is the graviton photoproduction amplitude, which can be diagrammatically represented as the sum of three diagrams:
\begin{equation}
i\mathcal{A}_{g\to\gamma}^{\text{tree}} \;=\;
\begin{tikzpicture}[baseline=(current bounding box.center)]
  % dotted incoming line
  \draw[dotted,line width=0.4pt] (-0.75,0) -- (0.75,0);
  % downward wavy line
  \draw[graviton]
       (0,0) -- (-0.75,-1);
  % downward wavy line
  \draw[decorate,decoration={zigzag,amplitude=1.2pt,segment length=6pt},line width=0.8pt]
       (0,0) -- (0.75,-1);
\end{tikzpicture}
\;+\;
\begin{tikzpicture}[baseline=(current bounding box.center)]
  % dotted incoming line
  \draw[dotted,line width=0.4pt] (-0.75,0) -- (-0.25,0);
  % short solid ledge
  \draw[line width=0.9pt] (-0.25,0) -- (0.25,0);
  % dotted outgoing line
  \draw[dotted,line width=0.4pt] (0.25,0) -- (0.75,0);
  % downward wavy line
  \draw[graviton]
       (-0.25,0) -- (-0.25,-1);
  \draw[decorate,decoration={snake,amplitude=1.2pt,segment length=6pt},line width=0.8pt]
       (0.25,0) -- (0.25,-1);
\end{tikzpicture}
\;+\;
\begin{tikzpicture}[baseline=(current bounding box.center)]
  \draw[dotted] (-0.75,0) -- (0.75,0);
  \draw[decorate,decoration={snake,amplitude=1.2pt,segment length=6pt},line width=0.8pt] (0,0) -- (0,-1);
  \draw[graviton]
       (-0.75,-1) -- (0,-1);
  \draw[decorate,decoration={snake,amplitude=1.2pt,segment length=6pt},line width=0.8pt]
       (0,-1) -- (0.75,-1);
\end{tikzpicture}.
\end{equation}
The full gauge-invariant scattering amplitude with general Wilson coefficients through $\mathcal{O}(\epsilon^2)$ is attached in Appendix~\ref{app:amplitude}. The bosonic field gauge invariance of our result has been checked by explicitly computing the Ward identity for both the graviton and photon. As we construct the graviton polarization tensor by the outer product of two polarization vectors, $\epsilon_h^{\mu\nu}=\epsilon_h^\mu\epsilon_h^\nu$, this requires replacing one polarization vector with the graviton momentum, $k_h^\mu\mathcal{A}_{\mu\nu}=0$. For the photon, $k_A^\mu\mathcal{A}_\mu=0$. The spin-gauge invariance is also explicitly checked by promoting the Lagrange multiplier with a scalar $\xi$, which we denote as the generalized $R_\xi$ gauge:
\begin{align}
\label{eq:spin-gauge}
    l_\mu&\rightarrow\frac{\xi}{p}\frac{Dp_\mu}{d\tau}.
\end{align}
Under this gauge, the Feynman rules are modified with $\xi$ dependence introduced (see Appendix~\ref{app:R_xi gauge}). Nevertheless, the amplitude should be independent of $\xi$ if it is truly spin-gauge invariant, which we have verified. Finally, in the spinless limit, the amplitude reduces to that of the graviton photoproduction by a spinless charged particle, formally identical to a RN black hole, worked out in~\cite{Ahmadiniaz_2020, PhysRevD.91.064008}. Our amplitude is consistent with these results.

The full scattering amplitude can be simplified in a well picked reference frame and appropriate residual gauge choices. In the rest frame of the black hole, $v=(1, 0, 0, 0)$. We can further choose to work in the gauge where the polarization vectors are purely spatial and transverse. These are conventionally denoted the radiation gauge for the photon and transverse traceless (TT) gauge for the graviton. As a result, any term with $(\epsilon_A\cdot v)$ or $(\epsilon_h\cdot v)$ vanishes. The simplified final scattering amplitude reads
\begin{align}
    &i\mathcal{A}_{g\to\gamma} = -ie\,\frac{\omega (k_A\cdot \epsilon_h)(\epsilon_h\cdot \epsilon_A)}{2(k_h\cdot k_A)} \notag\\
    &\quad + \frac{e}{2m}\Bigg[
        -\epsilon^{\epsilon_h\epsilon_A S v}(k_A\cdot \epsilon_h)
        + \epsilon^{\epsilon_h k_A S v}\frac{(k_A\cdot \epsilon_h)(k_h\cdot \epsilon_A)}{k_h\cdot k_A} \notag\\
    &\qquad
        - \epsilon^{k_h k_A S v}\frac{(\epsilon_h\cdot \epsilon_A)(k_A\cdot \epsilon_h)}{k_h\cdot k_A}
        - \epsilon^{\epsilon_A k_h S v}\frac{(k_A\cdot \epsilon_h)^2}{k_h\cdot k_A} \notag\\
    &\qquad
        + \epsilon^{k_A \epsilon_A S v}\frac{(k_A\cdot \epsilon_h)^2}{k_h\cdot k_A}
        + \epsilon^{\epsilon_h k_h S v}\frac{(k_A\cdot \epsilon_h)(k_h\cdot \epsilon_A)}{k_h\cdot k_A}
    \Bigg] \notag\\
    &\quad + \frac{ie}{4m^2}\Bigg[
        \frac{(S\cdot k_h)^2\,\omega (k_A\cdot \epsilon_h)(\epsilon_h\cdot \epsilon_A)}{k_h\cdot k_A} \notag\\
    &\qquad
        +\frac{(S\cdot k_A)^2\,\omega (k_A\cdot \epsilon_h)(\epsilon_h\cdot \epsilon_A)}{k_h\cdot k_A} \notag\\
    &\qquad
        +\frac{2(S\cdot k_h)(S\cdot k_A)\,\omega (k_A\cdot \epsilon_h)(\epsilon_h\cdot \epsilon_A)}{k_h\cdot k_A} \notag\\
    &\qquad 
        -4(S\cdot k_h)(S\cdot \epsilon_h)(\epsilon_h\cdot \epsilon_A)\omega\notag\\
    &\qquad
        +2(S\cdot \epsilon_h)^2\,\omega (k_h\cdot \epsilon_A)
        +2(S\cdot \epsilon_h)(S\cdot \epsilon_A)\,\omega (k_A\cdot \epsilon_h) \notag\\
    &\qquad
        -4(S\cdot \epsilon_h)(S\cdot k_A)\omega(\epsilon_h\cdot\epsilon_A)\notag\\
    &\qquad+\frac{2\,\epsilon^{k_A \epsilon_A S v}\,\epsilon^{k_h \epsilon_h S v}(k_A\cdot \epsilon_h)}{\omega}\Bigg],
\end{align}
where we have also plugged in the matched values of the Wilson coefficients for a KN black hole. The spin is now encoded in the spin vector $S^\mu=\frac{1}{2}\epsilon^{\mu\nu\rho\sigma}v_\nu S_{\rho\sigma}$, and we have used the shorthand notation $\epsilon^{\epsilon_Ak_ASv}=\epsilon^{\mu\nu\rho\sigma}\epsilon_{A\mu}k_{A\nu}S_\rho v_\sigma$. $\omega=k_h\cdot v=-k_A\cdot v$ is the graviton/photon energy in the black hole rest frame. The energy conserving delta function $2\pi\delta(k_h\cdot v+k_A\cdot v)$ is suppressed here in the expression, where we take all momenta to be incoming.

\subsection{Cross section}
\label{sec:xsec}
To fully understand the spin-induced corrections to the angular distribution of the scattering probability through $\mathcal{O}(S^2)$, we compute the differential cross section of the graviton photoproduction process. Starting from the scattering amplitude:
\begin{equation}
i\mathcal{A} = i (2\pi)\, \delta(E_f - E_i)\, \mathcal{M},
\end{equation}
the differential cross section simplifies to
\begin{align}
&\frac{d\sigma}{d\Omega}
= \frac{|\mathcal{M}|^2}{(4\pi)^2}\notag\\
&=\frac{1}{16\pi^2}\big(|\mathcal{M}_0|^2+2\text{Re}(\mathcal{M}_0^*\mathcal{M}_1)+2\text{Re}(\mathcal{M}_0^*\mathcal{M}_2)+|\mathcal{M}_1|^2\big),
\end{align}
where $\mathcal{M}_n$ denotes the matrix element at $\mathcal{O}(S^n)$. At the tree level, the classical differential cross section of plane wave scattering is identical to the quantum counterpart derived from the single particle state transition matrix, which we show explicitly in Appendix~\ref{app:xsec}. 
We continue to work in the rest frame of the black hole such that the spin is aligned with the z-axis. The incoming graviton is described by the azimuthal and polar angle $\alpha, \theta$ relative to the spin, while the direction of the outgoing photon is defined by the azimuthal and polar angle $\beta, \phi$. In Appendix~\ref{app:xsec} we show more details of how the polarizations and momenta are defined under this parameterization, with the geometry shown in Fig.~\ref{fig: geometry}. 

In a hypothetical measurement of the graviton photoproduction, the most relevant observable is likely the unpolarized differential cross section. We therefore first consider the case in which the incoming graviton polarization is unspecified and the outgoing photon polarization is not measured, corresponding to an average over initial graviton polarizations and a sum over final photon polarizations. In this unpolarized observable, the leading spin-dependent correction turns out to enter only at $\mathcal{O}(S^2)$, whereas for fixed-helicity channels the $\mathcal{O}(S)$ contribution is generally nonzero. Since the power counting in spin $S$ maps to the power counting in $\epsilon$ in the long wavelength regime, the $\mathcal{O}(S^2)$ is parametrically suppressed.

This suppression is further supported by the KN spin bound~\cite{Newman:1965my},
\begin{align}
S^2 \leq m^2\bigl(G^2m^2 - G e^2\bigr),
\end{align}
which implies that the spin-induced correction cannot be enhanced by an arbitrarily large spin. The combined effect of these two features suggests that spin corrections to the unpolarized differential cross section are indeed generically suppressed, and in the asymptotic long-wavelength limit the KN black hole behaves just as a RN black hole in terms of the graviton photoproduction.

\begin{figure*}[]
    \centering
    \subfigure[]{\includegraphics[width=0.9\columnwidth]{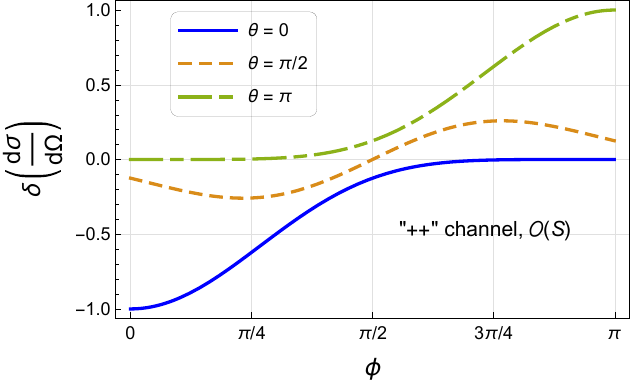}}
    \subfigure[]{\includegraphics[width=0.9\columnwidth]{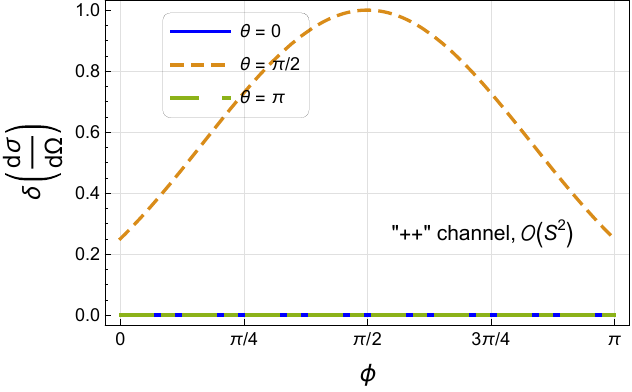}} 
    \subfigure[]{\includegraphics[width=0.9\columnwidth]{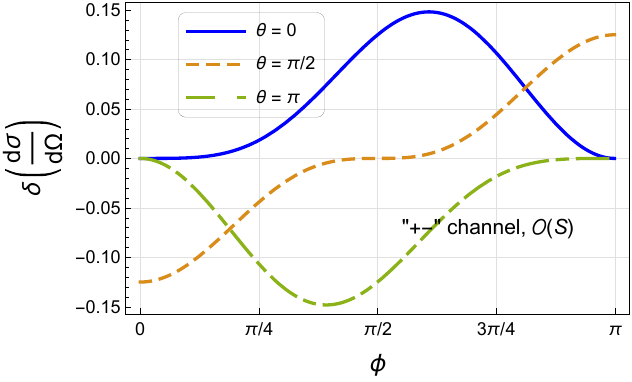}}
    \subfigure[]{\includegraphics[width=0.9\columnwidth]{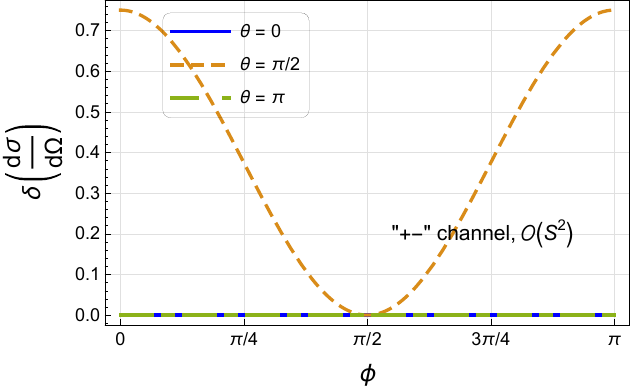}}
    \caption{The differential cross sections display the expected symmetry. The polar asymmetry introduced by the spin is manifest from the $\mathcal{O}(S)$ correction to the differential cross section. The $\mathcal{O}(S^2)$ corrections vanish when the incident graviton is either aligned or anti-aligned with the spin for individual helicity channels. Here we show spin induced correction to the differential cross section as a function of the polar angle of the outgoing photon momentum, $\phi$. A slice in the outgoing momentum sphere where the incident graviton and outgoing photon momenta are in the same plane is chosen. Three incident graviton directions are considered: the spin-aligned, the spin anti-aligned, and the spin-orthogonal. }
    \label{fig:xsec_line_plot}
\end{figure*}

As the KN black hole has a preferred direction due to the spin, the incoming graviton's direction also affects the scattering outcome. We emphasize three special cases: (1) when the graviton momentum aligns with the spin; (2) when the graviton momentum is orthogonal to the spin; (3) when the graviton momentum anti-aligns with the spin. The spin induced correction patterns to the full differential cross section are shown in Fig.~\ref{fig:xsec_line_plot}, where we take the incident and outgoing particles to be in the same plane as an example. The results on the full outgoing momentum sphere can be found in Fig.~\ref{fig:Unpolarized} of Appendix~\ref{app:xsec}. The overall differential cross section is larger in the forward direction, as expected in a scattering event. The spin induced corrections are symmetric between the spin-aligned and anti-aligned cases for the unpolarized scattering, which is consistent with the observation that the $\mathcal{O}(S)$ contributions vanish for unpolarized scattering. The $\mathcal{O}(S^2)$ corrections are necessarily polar symmetric since the cross section at this order is even under $S\to-S$.

The polarized helicity channels make the spin-induced asymmetry of the scattering process manifest, as shown in Fig.~\ref{fig:conserving} and Fig.~\ref{fig:flipping}. Unlike the unpolarized cross section, fixed-helicity differential cross sections generally receive contributions at both $\mathcal{O}(S)$ and $\mathcal{O}(S^2)$. These terms modify the angular dependence differently in the helicity-conserving and helicity-reversing channels, producing distinct scattering patterns.
\begin{figure}[t]
\centering
\resizebox{0.48\textwidth}{!}{%
\begin{tikzpicture}[
    line cap=round,
    line join=round,
    >=Latex,
    font=\Large,
    photon/.style={
        decorate,
        decoration={snake,amplitude=2.2pt,segment length=10pt},
        very thick
    },
    graviton/.style={
        decorate,
        decoration={coil,aspect=0.55,amplitude=2.0pt,segment length=4.5pt},
        very thick
    },
    mom/.style={-Latex, thick},
    thinmom/.style={-Latex, semithick},
    spinarrow/.style={-Latex, ultra thick},
    extend/.style={dashed, thin}
]

% -------------------------------------------------
% Black hole
% -------------------------------------------------
\def\RBH{0.2}
\coordinate (O) at (0,0);

\shade[ball color=black!85] (O) circle (\RBH);

% Spin vector along +z
\draw[spinarrow] (O) -- ++(0,1.2);
\node[right] at ($(O)+(0.15,2.65)$) {$m$, $e$, $\vec{S}$};

% -------------------------------------------------
% Incoming graviton and outgoing photon
% -------------------------------------------------
\coordinate (Gin)  at (-1.8,-3.6);
\coordinate (Ghit) at (0,0);
\coordinate (Phit) at (0,0);
\coordinate (Pout) at (3.8,0.55);

% Graviton: curly line
\draw[graviton] (Gin) -- (Ghit);

% Photon: wavy line
\draw[photon] (Phit) -- (Pout);

% Momentum arrows
\draw[mom]
  ($($(Gin)!0.28!(Ghit)$)+(0,-0.58)$) --
  ($($(Gin)!0.54!(Ghit)$)+(0,-0.58)$);

\draw[mom]
  ($($(Phit)!0.68!(Pout)$)+(0,0.28)$) --
  ($($(Phit)!0.94!(Pout)$)+(0,0.28)$);

% Text labels
\node[left]  at ($(O)+(3.8,4.0)$) {Kerr--Newman background};
\node[left]  at ($(Gin)+(-0.15,-0.45)$) {graviton};
\node[right] at ($(Pout)+(0.15,0.25)$) {photon};

\node[right] at ($(Gin)+(1.15,0.45)$) {$\vec{k}_h$};
\node[right] at ($(Pout)+(-0.85,0.95)$) {$\vec{k}_A$};

\node[right] at ($(O)+(0,-0.65)$) {$\alpha$};
\node[right] at ($(O)+(0.2,-1.25)$) {$\beta$};

\node[right] at ($(O)+(0.08,0.95)$) {$\theta$};
\node[right] at ($(O)+(0.2,1.65)$) {$\phi$};

\node[left] at ($(O)+(-0.1,3.2)$) {$z$};
\node[left] at ($(O)+(-3,-1)$) {$x$};
\node[left] at ($(O)+(3,-2.0)$) {$y$};

% -------------------------------------------------
% Reference rays for the two directions
% -------------------------------------------------
\draw[thinmom] (O) -- ++(-2.8,-1.55);
\draw[thinmom] (O) -- ++( 2.8,-1.55);
\draw[thinmom] (O) -- ++(0,3);

% -------------------------------------------------
% Angular guide arcs
% -------------------------------------------------
\draw[thin] ($(O)+(0,1.35)$) arc[start angle=90,end angle=63,radius=1.35];

\draw[thin] ($(O)+(2.28,0.33)$) arc[start angle=20,end angle=90,radius=2.42];

\draw[thin] ($(O)+(-0.85,-0.46)$) arc[start angle=210,end angle=395,radius=0.8865];

\draw[thin] ($(O)+(-1.35,-0.77)$) arc[start angle=210,end angle=355,radius=1.55];

% Projected guide lines
\draw[extend] (O) -- ++(1.2,2.4);
\draw[extend] (O) -- ++(1.2,0.95);
\draw[extend] (1.2,2.4) -- (1.2,0.95);
\draw[extend] (O) -- ++(3.8,-0.25);
\draw[extend] (Pout) -- (3.8,-0.25);

\end{tikzpicture}%
}
\caption{The geometry of the graviton photoproduction by scattering with a KN black hole. The definition of the parameters are labeled, where the spin is aligned with the z axis by choice. The black hole spin introduces a polar asymmetry that manifests in the spin dependent differential cross section. This asymmetry reduces in certain limits.}
\label{fig: geometry}
\end{figure}

By computing the spin corrections to the differential cross section as functions of the outgoing photon direction in both helicity-conserving and helicity-reversing channels, we find that spin introduces a transformation rule into the angular distribution: for the same incident graviton direction, the change in the spin correction at the same outgoing photon momentum corresponds to a flipped black hole spin between opposite helicity channels, for example ``++'' versus ``-- --'' and ``+ --'' versus ``-- +''.
\begin{align}
    \Delta|\mathcal{M}_{++}|^2&=\Delta|\mathcal{M}_{--}|^2 (S^\mu\to -S^\mu)\notag\\
    \Delta|\mathcal{M}_{+-}|^2&=\Delta|\mathcal{M}_{-+}|^2 (S^\mu\to -S^\mu)
\label{eq: xsec_symmetry}
\end{align}
This transformation rule is the root cause that the $\mathcal{O}(S)$ contribution to the unpolarized differential cross section vanishes, since a sum over helicity channels cancels out.

Another particularly notable cancellation occurs when the incident graviton is aligned or anti-aligned with the black hole spin. At $\mathcal{O}(S^2)$, the spin-induced correction to the differential cross section for fixed outgoing photon momentum also satisfies
\begin{equation}
\Delta |\mathcal{M}|^2(\phi,S)=\Delta |\mathcal{M}|^2(\phi,-S)
\end{equation}
within each individual helicity channel. In the aligned or anti-aligned configuration, the $\mathcal{O}(S^2)$ correction not only becomes azimuthally symmetric, but vanishes identically. At the level of the cross section, this arises from an exact cancellation between the quadratic contribution $|\mathcal{M}_1|^2$ and the interference term $2\mathrm{Re}\left(\mathcal{M}_0^*\mathcal{M}_2\right)$. Together with the universal absence of an $\mathcal{O}(S)$ correction to the unpolarized cross section, this implies that the black hole spin is invisible in the scattering process when the incident graviton momentum is parallel or antiparallel to the spin, unless the individual polarization channels are resolved. Spin effects become more pronounced as the incident momentum moves away from the spin axis, and are strongest when it is close to orthogonal to the spin.

\section{Discussions}
\label{sec:discussion}

In this work, we have presented the first explicit computation of the graviton photoproduction amplitude by a KN black hole in the low-energy regime, carried out systematically up to $\mathcal{O}(\epsilon^2)$. Our analysis identifies a low-energy sector of the KN EFT in which the mixed gravito-electromagnetic response is entirely determined by the background multipole structure of the KN solution to GR. By matching to the asymptotic gravitational and Maxwell fields, we demonstrate that spin-induced effects up to $\mathcal{O}(S^2)$ can be incorporated without directly solving the coupled Teukolsky system.

By working in the long-wavelength regime, this construction bypasses the long-standing technical obstacle posed by the nontrivial coupling of gravitational and electromagnetic perturbations in the KN background in a near-horizon treatment, while remaining fully controlled with the power counting. The resulting amplitude therefore provides a concrete and independent benchmark for future perturbative calculations, complementary to geometric-optics analyses in the high-frequency limit~\cite{PhysRevD.16.2915}. In addition, we have shown that the inclusion of electromagnetic interactions preserves the first-class constraint structure of the worldline theory. This provides a nontrivial consistency check of the formalism and supports its use as a general EFT framework for classical objects that are gravitating, spinning, and electrically charged simultaneously. Quantum corrections from graviton and photon loops can be incorporated in the current EFT framework, but are suppressed by separate quantum scales and lie beyond the classical KN response considered here. Extending to include the higher-order operators, non-conservative effects with mixed gravity-EM dynamics may be further explored. Such an extension would also make it possible to examine whether the mixed graviton–photon amplitudes display any remnant factorization structure, and to what extent they survive or break down in the classical spinning case, especially in comparison with the better-known relations for quantum amplitudes of finite-spin particles~\cite{PhysRevD.91.064008, Holstein_1_2006, Holstein_2_2006}. A careful exploration of these questions could help clarify which structures are specific to the quantum spinning particles and which persist in the classical charged particles with a non-vanishing angular momentum. Moreover, the simple structure of the operator basis for black holes might be more universal. In 4D, a KN black hole is shown to have no static polarizability when immersed in either an external EM field or a gravitational field~\cite{Charalambous_2021, Charalambous_2022}. If this translated to vanishing ``Love numbers'' for the mixed local operators, it would imply that the conservative sector of the graviton photoproduction by a 4D black hole is entirely fixed by its multipole moments, even beyond the order considered here.

Although astrophysical black holes are generally expected to be nearly electrically neutral because of efficient charge neutralization, several scenarios have been proposed in which primordial black holes (PBHs) could retain non-negligible charge. For example, it has been argued that for sufficiently light PBHs, Schwinger pair production may be suppressed, allowing the black hole to maintain electric charge over cosmological timescales~\cite{deFreitasPacheco:2024eij}. Another possibility is that PBHs carry charge under a hidden or dark $U(1)$ gauge symmetry, whose phenomenology can closely resemble that of ordinary electromagnetism~\cite{schuster2025primordialblackholescharge, Aly:2025dbx}. In such settings, the graviton--photon conversion process studied here could in principle have been active in the early Universe. While direct observability would likely require extreme conditions---including near-maximal charges, small black hole masses, and high number densities---the process nevertheless offers a theoretically clean probe of mixed electromagnetic--gravitational dynamics in strong-field environments.

From a phenomenological perspective, it is also interesting to ask whether mixed gravitational--electromagnetic couplings of the type captured by the present EFT could become relevant in astrophysical environments beyond the black-hole setting. Strongly magnetized neutron stars and magnetars provide one possible arena, since their large electromagnetic fields may enhance such effects. In compact binaries involving magnetized objects, these couplings could in principle induce additional contributions to the conservative dynamics, radiation, or waveform phasing beyond a purely gravitational description~\cite{Henry:2023guc, Henry:2023len, Zhang:2026vdd, Tang:2025jop}. While assessing their actual size would require a more realistic treatment of the ambient matter and magnetospheric physics, the EFT developed here offers the first step towards a systematic language for isolating and parameterizing such effects. It would therefore be interesting to explore whether any mixed gravity--electromagnetic signatures could become relevant for future observations.

\section{Acknowledgment}
The author is grateful to Maor Ben-Shahar and Irvin Mart\'{\i}nez-Rodr\'{\i}guez for useful discussions and valuable feedback. This work was supported by the Yale University fellowship. 

\onecolumngrid
\newpage

\appendix

\section{Equations of motion and first-class constraints}
\label{app:eomsandconstraint}
The constraints corresponding to both the reparameterization invariance and the gauge transformations in Eq.~(\ref{eq:gauge transformations}) must be first-class to be physically valid~\cite{Henneaux:1992ig, benshahar2023scatteringspinningcompactobjects}. These constraints are first-class if they are preserved at all times on the constraint surface: $\frac{DC}{d\tau}=0$ for $C=p^2-\mathcal{N}$ and $C=S^{\mu\nu}(\hat{p}_\nu+\Lambda_{0\nu})$, where $\mathcal{N}=m^2+\mathcal{L}_{n.m.}$. In this section we show that both constraints are indeed first-class with EM interactions included in the action, by working out the full equations of motion with Lagrange multipliers.

Starting with Eq.~(\ref{eq:master_action}), we vary the action with respect to $p_\mu, x_\mu, S_{\mu\nu}$ and $\Lambda^a_\mu$ respectively. For $p_\mu$ variation, we obtain
\begin{align}
\label{eq: p_eom}
    \frac{Dx^\mu}{d\tau}&=l\, p^\mu-\frac{l}{2}\frac{\partial \mathcal{N}}{\partial p_\mu}+\frac{1}{p}(\delta^\mu_\nu-\hat{p}^\mu\hat{p}_\nu)\cdot \Big(\frac{DS^{\nu\lambda}}{d\tau}\hat{p}_\lambda+2S^{\nu\lambda}\frac{D\hat{p}_\lambda}{d\tau}+l_\lambda S^{\lambda \nu}\Big).
\end{align}
Notice that setting spin to zero recovers the spinless equation of motion in a familiar form.
By varying with respect to $x_\mu$, we get
\begin{align}
\label{eq: x_eom}
    \frac{Dp_\mu}{d\tau}=\frac{1}{2}R_{\mu\nu}^{\, \, \, \, \rho\sigma}S_{\rho\sigma}\dot{x}^\nu+eF_{\mu\nu}\dot{x}^\nu+\frac{l}{2}D_\mu\mathcal{N}.
\end{align}
Here we have an additional Lorentz force term on top of the usual Mathisson–Papapetrou–Dixon equation, as expected. The remaining two equations take the same form as the case without the EM interaction~\cite{benshahar2023scatteringspinningcompactobjects}, and read
\begin{subequations}
\label{eq: S_and_Lambda_eoms}
\begin{align}
    \Omega^{\mu\nu}
+ 2 \frac{D \hat{p}^{[\mu}}{d\tau} \hat{p}^{\nu]}
- 2 \ell^{[\mu} \left( p^{\nu]} + \Lambda^{\nu]}_0 \right)
+ \ell N^{\mu\nu}
&= 0 \,, \\
\frac{D}{d\tau} S_{\mu\nu}
- 2 S_{[\mu|\rho} \Omega^\rho{}_{\nu]}
- 2 \ell_\rho S^\rho{}_{[\mu} \Lambda_{\nu]0}
&= 0 \,.
\end{align}
\end{subequations}

To check that $S^{\mu\nu}(\hat{p}_\nu+\Lambda_{0\nu})$ is first-class, we only need the two equations in Eq.~(\ref{eq: S_and_Lambda_eoms}). The pure gravity case has been explicitly shown in \cite{benshahar2023scatteringspinningcompactobjects}, and the modifications due to introducing EM interactions are masked in $\mathcal{N}$, which does not alter the formal derivation. By working on the constraint surface where $S^{\mu\nu}(\hat{p}_\nu+\Lambda_{0\nu})=0$, it can be shown that
\begin{align}
    \frac{D}{d\tau}S^{\mu\nu}(\hat{p}_\nu+\Lambda_{0\nu})&\approx S^{\mu\nu}\frac{D\hat{p}_\nu}{d\tau}+S_{\mu\rho}\Omega^{\rho}_{\,\,\nu}\hat{p}^\nu+l_\rho S^{\rho\mu}(\Lambda_0\cdot \hat{p}+1)\notag\\
    &\approx S_{\mu\rho}\hat{p}^\rho\frac{D\hat{p}^\nu}{d\tau}\hat{p}_\nu \notag\\
    &=0.
\end{align}
Here ``$\approx$'' means the equality holds on the constraint surface.

To check that $(p^2-\mathcal{N})=0$ is first-class, we start by expanding the covariant time derivative
\begin{align}
\label{eq: mass_shell_first_class}
    \frac{D}{d \tau}(p^2-\mathcal{N})&=2p^\mu\frac{Dp_\mu}{d\tau}-\dot{x}^\mu D_\mu\mathcal{N}-\frac{\partial \mathcal{N}}{\partial p^\mu}\frac{Dp^\mu}{d\tau}-\frac{\partial\mathcal{N}}{\partial S^{\mu\nu}}\frac{DS^{\mu\nu}}{d\tau}.
\end{align}
Making substitutions with Eq.~(\ref{eq: p_eom}), Eq.~(\ref{eq: x_eom}) and Eq.~(\ref{eq: S_and_Lambda_eoms}) in order, we rewrite the RHS of Eq.~(\ref{eq: mass_shell_first_class}) such that
\begin{align}
    \frac{D}{d \tau}(p^2-\mathcal{N})&=\frac{2\dot{x}^\mu}{l}\frac{Dp_\mu}{d\tau}-\dot{x}^\mu D_\mu\mathcal{N}-\frac{2}{l}\Big(\frac{DS^{\mu\nu}}{d\tau}\hat{p}_\nu+2S^{\mu\nu}\frac{D\hat{p}_\nu}{d\tau}+l_\nu S^{\nu\mu}\Big) \frac{D\hat{p}_\mu}{d\tau}-\frac{\partial \mathcal{N}}{\partial S^{\mu\nu}}\frac{DS^{\mu\nu}}{d\tau}\notag\\
    &=-\frac{2}{l}\Big(\frac{DS^{\mu\nu}}{d\tau}\hat{p}_\nu+l_\nu S^{\nu\mu}\Big)\frac{D\hat{p}_\mu}{d\tau}-\frac{\partial\mathcal{N}}{\partial S^{\mu\nu}}\frac{DS^{\mu\nu}}{d\tau}\notag\\
    &\approx-\frac{2l_\rho}{l}\Big(S^{\rho\mu}\Omega_{\mu\nu}\hat{p}^\nu+S^{\rho\mu}\frac{D\hat{p}_\mu}{d\tau}\Big)\notag\\
    &=-\frac{2l_\rho}{l}\Big(\frac{D}{d\tau}\Big[S^{\rho\mu}(\hat{p}_\mu+\Lambda_{\mu 0})\Big]-l_\mu S^{\mu\rho}(\Lambda_0\cdot \hat{p}+1)\Big)\notag\\
    &=-\frac{2l_\rho}{l}\frac{D}{d\tau}\Big[S^{\rho\mu}(\hat{p}_\mu+\Lambda_{\mu 0})\Big]\notag\\
    &\approx0.
\end{align}
In the last step we find that the covariant time derivative of the mass shell constraint naturally vanishes by the earlier observation that the SSC gauge fixing constraint is first-class, so the mass shell constraint is also first-class, as it should be. Importantly, the EM interaction does not spoil the first-class nature of the constraints and the Lorentz force term drops out of the expression by the antisymmetry of the Maxwell tensor.
\section{Vertex Feynman rules}
\label{app:feynman rules}
Here we document the full single graviton/photon emission vertex Feynman rules with contributions from the non-minimal operators up to $\mathcal{O}(\epsilon^2)$. 
\begin{align}
\begin{tikzpicture}[baseline=(current bounding box.center)]
  \draw[dotted] (-0.75,0) -- (0.75,0);
  \draw[graviton] (0,0) -- (0,-1);
\end{tikzpicture}
&= -\frac{i}{2} m (v \cdot h \cdot v)
   - \frac{1}{2} (v \cdot h \cdot S \cdot k)-\frac{ic_4}{4m}(k\cdot v)^2(S\cdot h\cdot S)-\frac{ic_4}{4m}(S\cdot k)^2(v\cdot h\cdot v)+\frac{ic_4}{2m}(k\cdot v)(S\cdot k)(S\cdot h\cdot v)\, .
\end{align}

\begin{align}
\begin{tikzpicture}[baseline=(current bounding box.center)]
  % dotted incoming line
  \draw[dotted,line width=0.4pt] (-0.75,0) -- (0,0);
  % short solid ledge
  \draw[line width=0.9pt] (0,0) -- (0.75,0);
  % downward wavy line
  \draw[graviton]
       (0,0) -- (0,-1);
\end{tikzpicture}
&=\frac{m}{2}\,(v \cdot h \cdot v)(k \cdot z)
- i\,(\pi \cdot h \cdot v) - \frac{i}{2}\,(v \cdot h \cdot S \cdot k)(z \cdot k) 
- \frac{1}{2}\,(v \cdot h \cdot s \cdot k) + \frac{i}{2}\,(z \cdot h \cdot S \cdot k)\,(k\cdot v)\notag \\[6pt]
&\quad + \frac{1}{2}\,(v \cdot h \cdot v)(v \cdot \lambda \cdot S \cdot k) - \frac{1}{2}\,(v \cdot h \cdot v)(v \cdot s \cdot k)-\frac{ic_4}{2m^2}(\pi \cdot h \cdot S\cdot S\cdot k)(k\cdot v)-\frac{ic_4}{2m^2}(v \cdot h \cdot S\cdot S\cdot k)(k\cdot \pi)\notag\\[6pt]
&\quad +\frac{ic_4}{2m^2}tr(h\cdot S\cdot S)(k\cdot v)(\pi\cdot k) +\frac{ic_4}{2m^2}(k\cdot S\cdot S\cdot k)(v\cdot h\cdot \pi)-\frac{c_4}{2m^2}(v \cdot h \cdot S)(k\cdot v)(S\cdot k)(k\cdot z) \notag\\[6pt]
&\quad +\frac{ic_4}{m^2}(v \cdot h \cdot S\cdot S\cdot k)(k\cdot v)(\pi\cdot v)-\frac{ic_4}{2m^2}tr(h\cdot S\cdot S)(k\cdot v)(k\cdot v)(\pi\cdot v) -\frac{ic_4}{2m^2}(k\cdot S\cdot S\cdot k)(v\cdot h\cdot v)(\pi\cdot v)\notag\\[6pt]
&\quad +\frac{c_4}{4m}(S \cdot h \cdot S)(k\cdot v)^2(k\cdot z) +\frac{c_4}{4m}(S\cdot k)^2(v\cdot h\cdot v)(k\cdot z)-\frac{ic_4}{2m}(v\cdot h\cdot S\cdot s\cdot k)(k\cdot v) \notag\\[6pt]
&\quad -\frac{ic_4}{2m}(k\cdot S\cdot s\cdot h\cdot v)(k\cdot v) +\frac{ic_4}{2m}tr(h\cdot S\cdot s)(k\cdot v)^2 +\frac{ic_4}{2m}(v\cdot h\cdot v)(k\cdot S\cdot s\cdot k)\notag\\[6pt]
\end{align}
\begin{align}
\begin{tikzpicture}[baseline=(current bounding box.center)]
  \draw[dotted] (-0.75,0) -- (0.75,0);
  \draw[decorate,decoration={snake,amplitude=1.2pt,segment length=6pt},line width=0.8pt] (0,0) -- (0,-1);
\end{tikzpicture}
= &-ie(v\cdot A)+\frac{c_2}{m}(k\cdot S\cdot A)-\frac{ic_3}{2m^2}(k\cdot S\cdot S\cdot k)(v\cdot A) +\frac{ic_3}{2m^2}(A\cdot S\cdot S\cdot k)(k\cdot v) .
\end{align}

\begin{align}
\begin{tikzpicture}[baseline=(current bounding box.center)]
  % dotted incoming line
  \draw[dotted,line width=0.4pt] (-0.75,0) -- (0,0);
  % short solid ledge
  \draw[line width=0.9pt] (0,0) -- (0.75,0);
  % downward wavy line
  \draw[decorate,decoration={snake,amplitude=1.2pt,segment length=6pt},line width=0.8pt]
       (0,0) -- (0,-1);
\end{tikzpicture}
&=e(v\cdot A)(k\cdot z)+e(z\cdot A)\omega +\frac{c_2}{m^2}(\pi\cdot S\cdot k)(A\cdot v) -\frac{c_2}{m^2}(\pi\cdot S\cdot A)(k\cdot v) +\frac{c_2}{m}(k\cdot s\cdot A) - \frac{c_2}{m}(k\cdot v)(v\cdot s\cdot A) \notag\\[6pt]
&\quad +\frac{c_2}{m}(A\cdot v)(v\cdot s\cdot k)+\frac{ic_2}{m}(k\cdot S\cdot A)(k\cdot z) -\frac{ic_3}{2m^3}(k\cdot S\cdot S\cdot k)(A\cdot \pi)
+\frac{ic_3}{2m^3}(k\cdot S\cdot S\cdot A)(k\cdot \pi) \notag\\[6pt]
&\quad +\frac{ic_3}{2m^3}(k\cdot S\cdot S\cdot k)(A\cdot v)(\pi\cdot v)-\frac{ic_3}{2m^3}(k\cdot S\cdot S\cdot A)(k\cdot v) (\pi\cdot v) +\frac{ic_3}{2m^3}(\pi\cdot S\cdot S\cdot k)(k\cdot v)(A\cdot v) \notag\\[6pt]
&\quad -\frac{ic_3}{2m^3}(\pi\cdot S\cdot S\cdot A)(k\cdot v)^2 -\frac{ic_3}{m^2}(k\cdot S\cdot s\cdot k)(A\cdot v) +\frac{ic_3}{2m^2}(k\cdot S\cdot s\cdot A)(k\cdot v) +\frac{ic_3}{2m^2}(A\cdot S\cdot s\cdot k)(k\cdot v) \notag\\[6pt]
&\quad -\frac{ic_3}{2m^2}(A\cdot S\cdot s\cdot v)(k\cdot v)^2 + \frac{ic_3}{2m^2}(k\cdot S\cdot s\cdot v)(k\cdot v)(A\cdot v) +\frac{c_3}{2m^2}(k\cdot S\cdot S\cdot k)(A\cdot v)(k\cdot z) \notag\\[6pt]
&\quad -\frac{c_3}{2m^2}(k\cdot S\cdot S\cdot A)(k\cdot v)(k\cdot z)\notag\\[6pt]
\end{align}

\begin{align}
&\begin{tikzpicture}[baseline=(current bounding box.center)]
  % dotted incoming line
  \draw[dotted,line width=0.4pt] (-0.75,0) -- (0.75,0);
  % downward wavy line
  \draw[graviton]
       (0,0) -- (-0.75,-1);
  % downward wavy line
  \draw[decorate,decoration={snake,amplitude=1.2pt,segment length=6pt},line width=0.8pt]
       (0,0) -- (0.75,-1);
\end{tikzpicture}
=\frac{c_2}{2m}\Big[-(k_A\cdot h\cdot S\cdot A)+(A\cdot h\cdot S\cdot k_A)-(v\cdot h\cdot S\cdot k_A)(A\cdot v)+(v\cdot h\cdot S\cdot A)(k_A\cdot v)\Big]\notag\\[6pt]
&\quad +\frac{ic_3}{2m^2}(k_A\cdot h\cdot S\cdot S\cdot k_A)(A\cdot v) +\frac{ic_3}{4m^2}(k_A\cdot h\cdot S\cdot S\cdot A)(k_h\cdot v)+\frac{ic_3}{4m^2}(k_A\cdot S\cdot S\cdot h\cdot A)(k_h\cdot v)\notag\\[6pt]
&\quad +\frac{ic_3}{2m^2}(k_h\cdot S\cdot S\cdot h\cdot k_A)(A\cdot v) -\frac{ic_3}{4m^2}tr(h\cdot S\cdot S)(k_h\cdot k_A)(A\cdot v) +\frac{ic_3}{2m^2}(k_h\cdot S\cdot S\cdot h\cdot A)(k_h\cdot v) \notag\\[6pt]
&\quad -\frac{ic_3}{4m^2}tr(h\cdot S\cdot S)(k_h\cdot A)(k_h\cdot v) +\frac{ic_3}{4m^2}(k_h\cdot S\cdot S\cdot k_A)(A\cdot h\cdot v) +\frac{ic_3}{4m^2}(A\cdot h\cdot S\cdot S\cdot k_A)(k_h\cdot v) \notag\\[6pt]
&\quad -\frac{ic_3}{4m^2}(v\cdot h\cdot S\cdot S\cdot k_A)(k_h\cdot A) -\frac{ic_3}{4m^2}(k_h\cdot S\cdot S\cdot A)(k_A\cdot h\cdot v) -\frac{ic_3}{4m^2}(k_A\cdot h\cdot S\cdot S\cdot A)(k_h\cdot v) \notag\\[6pt]
&\quad +\frac{ic_3}{4m^2}(v\cdot h\cdot S\cdot S\cdot A)(k_h\cdot k_A)+\frac{ic_3}{2m^2}(A\cdot h\cdot v)(k_A\cdot S\cdot S\cdot k_A)-\frac{ic_3}{2m^2}(k_A\cdot h\cdot v)(k_A\cdot S\cdot S\cdot A)\notag\\[6pt]
&\quad + \frac{ic_3}{4m^2} (v\cdot h\cdot S\cdot S\cdot k_A)(k_h\cdot v) + \frac{ic_3}{4m^2} (v\cdot h\cdot S\cdot S\cdot A)(k_h\cdot v)^2-\frac{ic_3}{4m^2}(k_A\cdot S\cdot S\cdot k_A)(v\cdot h\cdot v)(A\cdot v)\notag\\[6pt]
&\quad -\frac{ic_3}{4m^2}(k_A\cdot S\cdot S\cdot A)(v\cdot h\cdot v)(k_h\cdot v)
\end{align}

\section{Full gauge-invariant scattering amplitudes}
\label{app:amplitude}
The full gauge-invariant scattering amplitudes are as follows. We have suppressed $\kappa$ as a common factor for the $g\to\gamma$ amplitudes, and suppressed the energy conserving $2\pi\delta(k_1\cdot v+k_2\cdot v)$ in all expressions.

\begin{align}
    i\mathcal{A}_0^{g\to\gamma} &= -ie\Bigg[
    \frac{k_h\cdot k_A}{2\omega^2}(\epsilon_h\cdot v)^2(\epsilon_A\cdot v)
    -\frac{(k_A\cdot \epsilon_h)(\epsilon_h\cdot v)(\epsilon_A\cdot v)}{\omega}
    +\frac{(k_h\cdot \epsilon_A)(\epsilon_h\cdot v)^2}{2\omega}\notag\\
    &\quad
    -\frac{(\epsilon_h\cdot v)(\epsilon_h\cdot \epsilon_A)}{2} 
    +\frac{(k_A\cdot \epsilon_h)^2(\epsilon_A\cdot v)}{2(k_h\cdot k_A)}
    -\frac{(k_A\cdot \epsilon_h)(k_h\cdot \epsilon_A)(\epsilon_h\cdot v)}{2(k_h\cdot k_A)}
    +\frac{(k_A\cdot \epsilon_h)(\epsilon_h\cdot \epsilon_A)\omega}{2(k_h\cdot k_A)}
    \Bigg],
\end{align}
\begin{align}
    i\mathcal{A}_1^{g\to\gamma}&=-\frac{e}{2m}\Bigg[\epsilon^{\epsilon_hk_hSv}\bigg(\frac{(\epsilon_h\cdot v)(\epsilon_A\cdot v)(k_h\cdot k_A)}{\omega^2}+\frac{(\epsilon_h\cdot v)(k_h\cdot \epsilon_A)}{\omega}-\frac{(\epsilon_h\cdot v)(k_A\cdot \epsilon_h)}{\omega}-(\epsilon_h\cdot \epsilon_A)\bigg)\Bigg]\notag\\
    &-\frac{c_2}{2m}\Bigg[\epsilon^{\epsilon_h\epsilon_ASv}\Bigg((k_A\cdot \epsilon_h)-\frac{(\epsilon_h\cdot v)(k_h\cdot k_A)}{\omega}-\omega(\epsilon_h\cdot v)\Bigg)+\epsilon^{k_h\epsilon_hSv}\Bigg(\frac{(k_A\cdot \epsilon_h)(k_h\cdot \epsilon_A)}{k_h\cdot k_A}-(\epsilon_h\cdot \epsilon_A)\Bigg)\notag\\
    &+\epsilon^{k_hk_ASv}\Bigg(\frac{(\epsilon_h\cdot \epsilon_A)(k_A\cdot \epsilon_h)}{k_h\cdot k_A}-\frac{(\epsilon_h\cdot v)(\epsilon_h\cdot \epsilon_A)}{\omega}\Bigg)+\epsilon^{\epsilon_Ak_hSv}\Bigg(\frac{(k_A\cdot \epsilon_h)^2}{k_h\cdot k_A}-\frac{(\epsilon_h\cdot v)(k_A\cdot \epsilon_h)}{\omega}\Bigg)\notag\\
    &+\epsilon^{\epsilon_h k_A Sv}\Bigg(\frac{(\epsilon_h\cdot v)(k_h\cdot \epsilon_A)}{\omega}-\frac{(k_A\cdot \epsilon_h)(k_h\cdot \epsilon_A)}{k_h\cdot k_A}-(\epsilon_h\cdot v)(\epsilon_A\cdot v)\Bigg)\notag\\ &+\epsilon^{k_A\epsilon_ASv}\Bigg(\frac{2(\epsilon_h\cdot v)(k_A\cdot \epsilon_h)}{\omega}-\frac{(\epsilon_h\cdot v)^2(k_h\cdot k_A)}{\omega^2}-\frac{(k_A\cdot \epsilon_h)^2}{k_h\cdot k_A}\Bigg)\Bigg],
\end{align}
\begin{align}
    i\mathcal{A}_2^{g\to\gamma}&=-\frac{ic_3}{4m^2}\Bigg[(S\cdot k_h)^2\Bigg(-(\epsilon_h\cdot v)(\epsilon_h\cdot \epsilon_A)-\frac{(k_A\cdot \epsilon_h)^2(\epsilon_A\cdot v)}{k_h\cdot k_A}+\frac{(k_A\cdot \epsilon_h)(k_h\cdot \epsilon_A)(\epsilon_h\cdot v)}{k_h\cdot k_A}-\frac{\omega(k_A\cdot \epsilon_h)(\epsilon_h\cdot \epsilon_A)}{k_h\cdot k_A}\Bigg)\notag\\
    &+(S\cdot k_A)^2\Bigg((\epsilon_h\cdot v)(\epsilon_h\cdot \epsilon_A)-\frac{(k_A\cdot \epsilon_h)^2(\epsilon_A\cdot v)}{k_h\cdot k_A}+\frac{(k_A\cdot \epsilon_h)(k_h\cdot \epsilon_A)(\epsilon_h\cdot v)}{k_h\cdot k_A}-\frac{\omega(k_A\cdot \epsilon_h)(\epsilon_h\cdot \epsilon_A)}{k_h\cdot k_A}\notag\\
    &-\frac{(\epsilon_h\cdot v)^2(k_h\cdot \epsilon_A)}{\omega}-\frac{(\epsilon_h\cdot v)^2(\epsilon_A\cdot v)(k_h\cdot k_A)}{\omega^2}+\frac{2(\epsilon_h\cdot v)(k_A\cdot \epsilon_h)(\epsilon_A\cdot v)}{\omega}\Bigg)\notag\\
    &+2(S\cdot k_h)(S\cdot k_A)\Bigg(-\frac{(k_A\cdot \epsilon_h)^2(\epsilon_A\cdot v)}{k_h\cdot k_A}+\frac{(k_A\cdot \epsilon_h)(k_h\cdot \epsilon_A)(\epsilon_h\cdot v)}{k_h\cdot k_A}-\frac{\omega(k_A\cdot \epsilon_h)(\epsilon_h\cdot \epsilon_A)}{k_h\cdot k_A}+\frac{(k_A\cdot \epsilon_h)(\epsilon_h\cdot v)(\epsilon_A\cdot v)}{\omega}\Bigg)\notag\\
    &+2(S\cdot \epsilon_h)(S\cdot k_A)\Bigg(+(k_A\cdot\epsilon_h)(\epsilon_A\cdot v)+\omega(\epsilon_h\cdot \epsilon_A)-(k_h\cdot \epsilon_A)(\epsilon_h\cdot v)-\frac{(k_h\cdot k_A)(\epsilon_h\cdot v)(\epsilon_A\cdot v)}{\omega}\Bigg)\notag\\
    &+2(S\cdot \epsilon_h)(S\cdot k_h)\Bigg((k_A\cdot \epsilon_h)(\epsilon_A\cdot v)+(\epsilon_h\cdot \epsilon_A)\omega\Bigg)-(S\cdot \epsilon_h)^2\Bigg((k_h\cdot k_A)(\epsilon_A\cdot v)+(k_h\cdot \epsilon_A)\omega\Bigg)\Bigg]\notag\\
    &+\frac{iec_4}{2m^2}\Bigg[(S\cdot k_h)^2\Bigg(\frac{(\epsilon_h\cdot v)(\epsilon_A\cdot v)(k_A\cdot \epsilon_h)}{\omega}+(\epsilon_h\cdot v)(\epsilon_h\cdot \epsilon_A)-\frac{(\epsilon_h\cdot v)^2(\epsilon_A\cdot v)(k_h\cdot k_A)}{2\omega^2}-\frac{(\epsilon_h\cdot v)(k_h\cdot \epsilon_A)}{2\omega}\Bigg)\notag\\
    &-(S\cdot k_h)(S\cdot \epsilon_h)\Bigg((k_A\cdot \epsilon_h)(\epsilon_A\cdot v)+\omega(\epsilon_h\cdot \epsilon_A)\Bigg)+(S\cdot \epsilon_h)^2\Bigg(\frac{(\epsilon_A\cdot v)(k_h\cdot k_A)}{2}+\frac{\omega(k_h\cdot \epsilon_A)}{2}\Bigg)\Bigg]\notag\\
    &+\frac{ic_2c_4}{2m^2}\Bigg[+(S\cdot k_A)(S\cdot \epsilon_h)(\epsilon_h\cdot v)(k_h\cdot\epsilon_A)-(S\cdot \epsilon_h)(S\cdot\epsilon_A)(\epsilon_h\cdot v)(k_h\cdot k_A)+(S\cdot k_h)(S\cdot k_A)(\epsilon_h\cdot v)(\epsilon_h\cdot \epsilon_A)\notag\\
    &-(S\cdot k_h)(S\cdot \epsilon_A)(\epsilon_h\cdot v)(k_A\cdot \epsilon_h)-(S\cdot \epsilon_h)(S\cdot k_A)\omega(\epsilon_h\cdot\epsilon_A)+(S\cdot \epsilon_h)(S\cdot \epsilon_A)\omega(k_A\cdot \epsilon_h)\notag\\
    &-\frac{(S\cdot k_h)(S\cdot k_A)(\epsilon_h\cdot v)^2(k_h\cdot \epsilon_A)}{\omega}+\frac{(S\cdot k_h)(S\cdot \epsilon_A)(\epsilon_h\cdot v)^2(k_h\cdot k_A)}{\omega}\Bigg]\notag\\
    &-\frac{ie}{2m^2}\Bigg[-\frac{(\epsilon_h\cdot v)(\epsilon_A\cdot v)\epsilon^{\epsilon_hk_hSv}\epsilon^{k_hk_ASv}}{\omega}-(\epsilon_h\cdot v)\epsilon^{\epsilon_hk_hSv}\epsilon^{k_h\epsilon_ASv}+(\epsilon_A\cdot v)\epsilon^{\epsilon_hk_hSv}\epsilon^{\epsilon_hk_ASv}+\omega\epsilon^{\epsilon_hk_hSv}\epsilon^{\epsilon_h\epsilon_ASv}\Bigg]\notag\\
    &-\frac{ic_2}{2m^2}\Bigg[\epsilon^{k_A\epsilon_ASv}\epsilon^{k_h\epsilon_hSv}\Bigg(\frac{(k_h\cdot k_A)(\epsilon_h\cdot v)}{\omega^2}-\frac{(k_A\cdot \epsilon_h)}{\omega}\Bigg)-\epsilon^{k_h\epsilon_A S v}\epsilon^{k_h\epsilon_hS v}(\epsilon_h\cdot v)+\epsilon^{k_h\epsilon_h S v}\epsilon^{\epsilon_h\epsilon_AS v}\omega\notag\\
    &-\epsilon^{k_h\epsilon_h S v}\epsilon^{k_h k_A S v}\frac{(\epsilon_h\cdot v)(\epsilon_A\cdot v)}{\omega}+\epsilon^{k_h\epsilon_hS v}\epsilon^{\epsilon_h k_A S v}(\epsilon_A\cdot v)\Bigg].
\end{align}

We check the Ward identity for both the graviton and photon, replacing a polarization vector with the external momentum of the corresponding external particle, and the resulting longitudinal modes are consistently zero. The gauge invariance should hold regardless of what the Wilson coefficients are, so each sector proportional to a particular combination of the Wilson coefficients should obey the Ward identity itself. For example, for the term proportional to $c_2$ in the $g\to\gamma$ scattering amplitude, $\epsilon^{k_hk_hSv}=0$ by anti-symmetry, so $k_h^\mu \mathcal{A}_{\mu\nu}\propto\frac{(k_h\cdot k_A)(k_h\cdot v)}{\omega^2}-\frac{k_A\cdot k_h}{\omega}=0$, hence the graviton Ward identity is preserved. Similarly, by anti-symmetry of $\epsilon^{k_Ak_ASv}$, the photon Ward identity holds. 

Furthermore, it may appear that the amplitude is organized as powers of $\omega S/m$ instead of $\omega m$. Fundamentally, the expansion should still be understood as a long wavelength expansion. This is because the black hole spin $S$ is not dimensionless. For KN black holes, the dimensionless spin parameter is $\chi_s=S/(Gm^2)$, which makes $\omega S/m=\chi_s G\, \omega m\sim \omega r_s$.

\section{Spin-gauge invariance}
\label{app:R_xi gauge}
On top of the bosonic field gauge invariance, the scattering amplitude should also be SSC spin-gauge invariant, which can be checked by adopting a generalized gauge parameterized by a scalar $\xi$ for the Lagrange multiplier $l_a$ as in Equation~(\ref{eq:master_action}). The propagators in this generalized $R_{\xi}$ gauge are as follows~\cite{benshahar2023scatteringspinningcompactobjects}:
\begin{subequations}
\begin{align}
\langle z^\mu(-\omega) z^\nu(\omega) \rangle 
&= - i \frac{1}{m\omega^2} \eta^{\mu\nu} 
   - \frac{1}{m^2 \omega} S^{\mu\nu} , \label{eq:a}\\[6pt]
\langle \pi^\mu(-\omega) z^\nu(\omega) \rangle 
&= - \frac{1}{\omega} \eta^{\mu\nu} , \label{eq:b}\\[6pt]
\langle s_{\mu\nu}(-\omega) s_{\rho\sigma}(\omega) \rangle 
&= - \frac{2}{\omega} 
   \big( \eta_{\nu[\sigma} S_{\rho]\mu} 
       - \eta_{\mu[\sigma} S_{\rho]\nu} \big) , \label{eq:d}\\[6pt]
\langle s_{\mu\nu}(-\omega) \lambda_{\rho\sigma}(\omega) \rangle 
&= \frac{2}{\omega} \, \eta_{\mu[\rho} \eta_{\sigma]\nu} , \label{eq:e}\\[6pt]
\langle \lambda^{\mu\nu}(-\omega) z_\rho(\omega) \rangle 
&= -(2\xi-1)\frac{2}{m\omega} \, v^{[\mu} \delta^{\nu]}_{\rho} , \label{eq:f}\\[6pt]
\langle z^\mu(-\omega) s^{\nu\rho}(\omega) \rangle
&= (1-\xi)\frac{2}{m\omega} S^{\mu[\nu}v^{\rho]} .
\label{eq:g}
\end{align}
\end{subequations}

The vertex rules also carry the $\xi$-dependence. Concretely, the minimal worldline perturbing graviton vertex becomes
\begin{align}
\begin{tikzpicture}[baseline=(current bounding box.center)]
  % dotted incoming line
  \draw[dotted,line width=0.4pt] (-0.75,0) -- (0,0);
  % short solid ledge
  \draw[line width=0.9pt] (0,0) -- (0.75,0);
  % downward wavy line
  \draw[graviton]
       (0,0) -- (0,-1);
\end{tikzpicture}
&=\frac{m}{2}\,(v \cdot h \cdot v)(k \cdot z)
- i\,(\pi \cdot h \cdot v) - \frac{i}{2}\,(v \cdot h \cdot S \cdot k)(z \cdot k) 
- \frac{1}{2}\,(v \cdot h \cdot s \cdot k) + \frac{i}{2}\,(z \cdot h \cdot S \cdot k)\,(k\cdot v)\notag \\[6pt]
&\quad + \frac{\xi}{2}\,(v \cdot h \cdot v)(v \cdot \lambda \cdot S \cdot k) + \frac{1-\xi}{2m}(v\cdot h\cdot v)(\pi\cdot S\cdot k)+\frac{1-2\xi}{2}\,(v \cdot h \cdot v)(v \cdot s \cdot k)\\[6pt]
&\quad + \frac{1-\xi}{2m}\omega (\pi\cdot S\cdot h\cdot v).
\end{align}
We explicitly check that the $\xi$-dependent pieces cancel out in the final scattering amplitude.

\section{Cross Section}
\label{app:xsec}
Here we explicitly show that the tree level quantum cross section computed in the main text has the same expression as the classical scattering cross section with an incoming gravitational plane wave and an outgoing electromagnetic plane wave. Starting from the full action (Eq.~\ref{eq:pathintegral}), we integrate out the worldline degrees of freedom to obtain an effective action
\begin{align}
    S_{\rm eff}=S_{\rm bulk}+S_{\rm gf}+S_{\rm int},
\end{align}
where 
\begin{align}
    S_{\rm int}=\int d^4x\int d^4y \,A_\mu(x)V^{\mu|\alpha\beta}(x, y)h_{\alpha\beta}(y)
\end{align}
is the effective interaction vertex that contains all the tree level diagrams with an effective kernel $V^{\mu|\alpha\beta}(x, y)$, which depends on the background parameters $m, e, S$. Including up to the quadratic(kinetic) bulk field terms and applying our gauge choices, we arrive at
\begin{align}
    S_{\rm eff}=-\frac{1}{2}\int d^4x A_\mu \Box A^{\mu}-\frac{1}{2}h_{\mu\nu}\Box P^{\mu\nu, \alpha\beta}_hh_{\alpha\beta}+\int d^4x\int d^4y \, A_\mu(x)V^{\mu|\alpha\beta}(x, y)h_{\alpha\beta}(y).
\end{align}
Varying with respect to the electromagnetic field $A_{\mu}$, we obtain the classical equation of motion for the electromagnetic field
\begin{align}
    \Box A^\mu=\int d^4y\,  V^{\mu|\alpha\beta}(x, y)h_{\alpha\beta}(y).
\end{align}
For an incident gravitational plane wave
\begin{align}
    h_{\alpha\beta}(y)&=h_0\epsilon_{\alpha\beta}(k)e^{-ik\cdot y},\notag\\
    k=&(\omega,\vec{k}), k^2=0,
\end{align}
it follows that
\begin{align}
    \Box A^\mu&=h_0\int d^4y\,  V^{\mu|\alpha\beta}(x, y)\epsilon_{\alpha\beta}(y)e^{-ik\cdot y}\notag\\
    &=J^\mu(x).
\end{align}
Fourier transforming to the $\omega$-space, this becomes
\begin{align}
    (\nabla^2+\omega'^2)A^\mu(\omega',\vec{x})=-J^\mu(\omega',\vec{x}).
\end{align}
Working in the far zone limit, the solution is~\cite{Jackson:1998nia}
\begin{align}
    A^{\mu}(\omega', \vec{x})&=\frac{e^{i\omega' r}}{4\pi r}\int d^3\vec{x}' e^{-i\vec{k}'\cdot \vec{x}'} J^{\mu}(\omega', \vec{x}').
\end{align}
Hence
\begin{align}
    A(t, \vec{x})&=\int \frac{d\omega'}{2\pi} \frac{e^{-i\omega't+i\omega'r}}{4\pi r}\int d^3\vec{x}' e^{-i\vec{k}'\cdot \vec{x}'}\int dt' e^{i\omega't'} J^{\mu}(t', \vec{x}')\epsilon_{\mu}(k')\notag\\
    &=\frac{h_0}{4\pi r}\int \frac{d\omega'}{2\pi} e^{-i\omega't+i\omega'r}\epsilon_{\mu}(k')\int d^4x' \int d^4y e^{-ik'\cdot x'}V^{\mu|\alpha\beta}(x, y)e^{-ik\cdot y}\epsilon_{\alpha\beta}(k).
    \label{eq:EM_outgoing_classical}
\end{align}
Meanwhile, the quantum transition matrix element is computed with
\begin{align}
    i\mathcal{A}&=\langle\gamma(k')|iS_{\rm int}|g(k)\rangle\notag\\
    &=i\int d^4x \int d^4y\, \epsilon^*(k')e^{ik'\cdot x}V^{\mu|\alpha\beta}(x, y)\epsilon_{\alpha\beta}(k)e^{-ik\cdot y}\notag\\
    &=i2\pi\delta(\omega-\omega')\mathcal{M},
\end{align}
as we have derived. Plugging back into Eq.~(\ref{eq:EM_outgoing_classical}) and integrating, the scattered electromagnetic wave takes the form:
\begin{align}
    A(t, \vec{x})=\frac{h_0}{4\pi r}e^{-i\omega t+i\omega r} \mathcal{M}.
\end{align}
The energy flux per solid angle is thus
\begin{align}
    dP_{\rm EM}/d\Omega&=r^2 \omega^2h_0^2\frac{|\mathcal{M}|^2}{(4\pi r)^2}\notag\\
    &=\omega^2h_0^2\frac{|\mathcal{M}|^2}{(4\pi)^2}.
\end{align}
The incoming gravitational wave energy flux is 
\begin{align}
    P_{\rm GW}=\omega^2h_0^2.
\end{align}
Together, the classical differential cross section of the plane wave scattering evaluates to
\begin{align}
    \frac{d\sigma_{\rm cl}}{d\Omega}&=\frac{dP_{\rm EM}/d\Omega}{P_{\rm GW}}\notag\\
    &=\frac{|\mathcal{M}|^2}{(4\pi)^2},
\end{align}
which is formally identical to the quantum scattering differential cross section we have derived.

In calculating the cross section, we define the graviton/photon azimuthal angle $\alpha/\beta$ and the polar angle $\theta/\phi$, with the notations reserved for graviton and photon in order, as illustrated in Fig.~\ref{fig: geometry}. The polar angle is defined such that 0 corresponds to the positive z-axis. The spin is chosen to be aligned with the positive z-axis. The momentum and polarization of the graviton and photon can thus be expressed as 
\begin{align}
    k_h&=\omega(1, \sin\theta\cos\alpha, \sin\theta\sin\alpha, \cos\theta),\\
    k_A&=-\omega(1, \sin\phi\cos\beta, \sin\phi\sin\beta, \cos\phi),\\
    \epsilon_h^\mu(\lambda=\pm1)&=\frac{1}{\sqrt{2}}(0, \cos\theta\cos\alpha+i\lambda\sin\theta\cos\alpha, \cos\theta\sin\alpha+i\lambda \sin\theta\sin\alpha, -\sin\theta-i\lambda \cos\theta),\\
    \epsilon_A^\mu(\lambda=\pm1)&=\frac{1}{\sqrt{2}}(0, \cos\phi\cos\beta-i\lambda\sin\phi\cos\beta, \cos\phi\sin\beta-i\lambda \sin\phi\sin\beta, -\sin\phi+i\lambda \cos\phi),
\end{align}
where we use opposite signs for the momenta and polarization vectors of the graviton and photon to be consistent with the all-incoming convention in our amplitudes. In the rest frame where the spin is aligned with the z axis, $S^\mu=(0,0,0,S)$, $v=(1,0,0,0)$. In this frame, our conventional expression also simplifies $\epsilon^{abSv}=\vec{a}\cdot(\vec{b}\times\vec{S})$. Fig.~\ref{fig: geometry} illustrates the geometry of the scattering process.

We explicitly calculate the differential cross section for the cases where the incoming graviton is aligned, anti-aligned, or orthogonal to the spin, where we set the overall scale factor $S\omega/m$ to unity for presentational purposes.

\begin{figure*}[h]
    \centering
    \subfigure[$\alpha=0, \theta=\pi/3$]{\includegraphics[width=0.32\columnwidth]{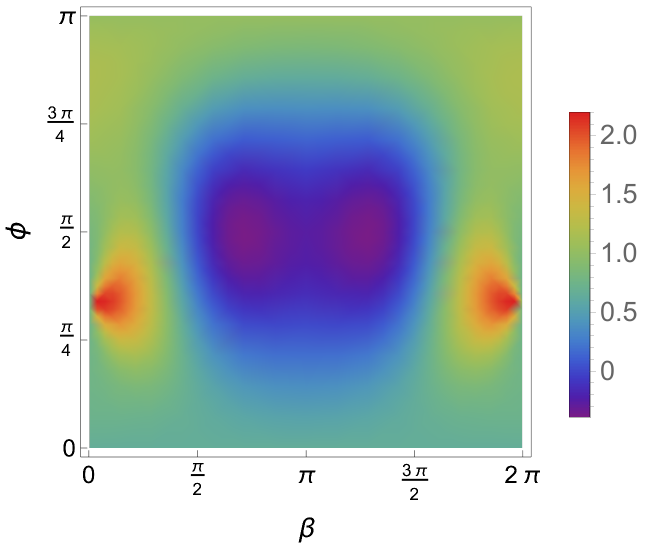}}
    \subfigure[$\alpha=0, \theta=\pi/2$]{\includegraphics[width=0.31\columnwidth]{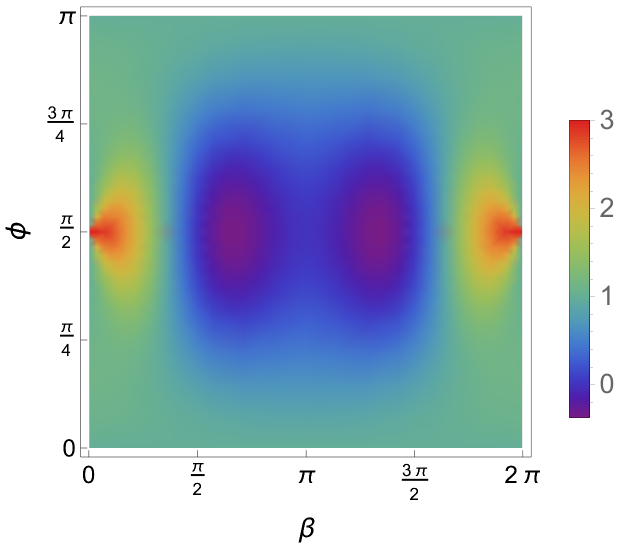}} \subfigure[$\alpha=0, \theta=2\pi/3$]{\includegraphics[width=0.32\columnwidth]{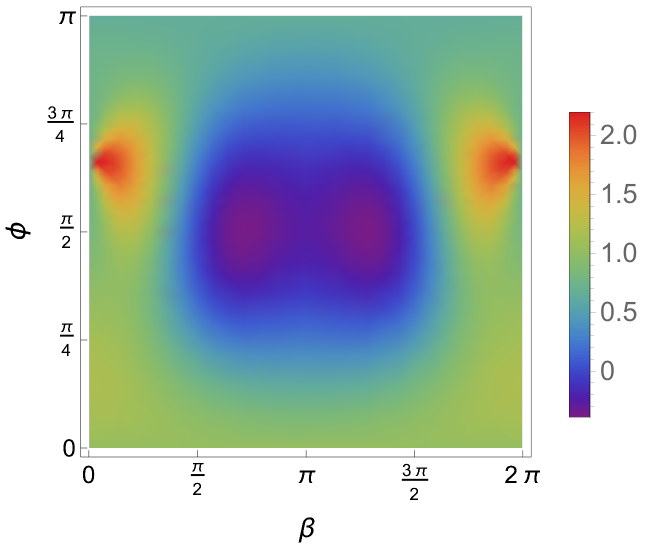}}
    \caption{The spin correction in the unpolarized differential cross sections of graviton photoproduction is symmetric between the directions aligned and anti-aligned with the spin. For incoming graviton-spin alignment/anti-alignment, the spin dependence drops out through $\mathcal{O}(S^2)$. Here we present the spin corrections for (1)$\alpha=0, \theta=\pi/3$, (2)$\alpha=0, \theta=\pi/2$, (3)$\alpha=0, \theta=2\pi/3$.}
    \label{fig:Unpolarized}
\end{figure*}

Fig.~\ref{fig:Unpolarized} shows the spin corrections for three representative incident directions for the unpolarized scattering. The unpolarized scattering differential cross section receives spin correction starting at $\mathcal{O}(S^2)$. When the incident graviton's momentum is aligned or anti-aligned with the spin, the $\mathcal{O}(S^2)$ contribution also vanishes. For other incoming graviton directions, the spin dependence enters. The polar symmetry is not broken by the presence of spin in the unpolarized scattering. In other words, the spin induced correction pattern to the differential cross section remains the same when the spin direction is flipped.

\begin{figure*}[h]
    \centering
    \subfigure[$\alpha=0, \theta=0$]{\includegraphics[width=0.325\columnwidth]{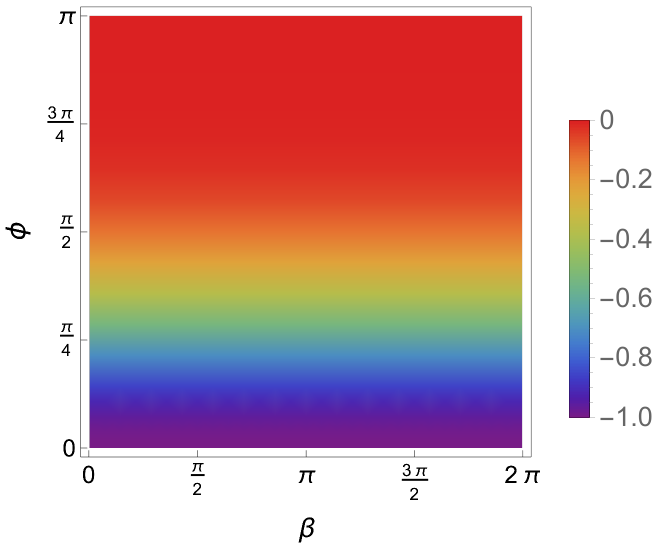}}
    \subfigure[$\alpha=0, \theta=\pi/3$]{\includegraphics[width=0.32\columnwidth]{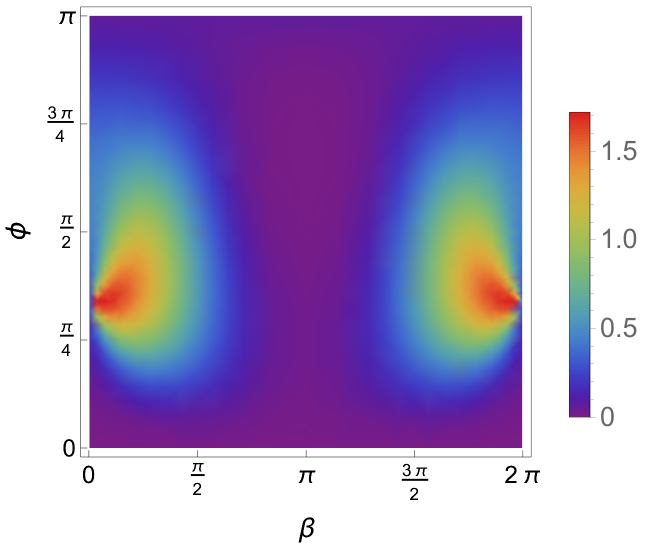}} \subfigure[$\alpha=0, \theta=\pi/2$]{\includegraphics[width=0.32\columnwidth]{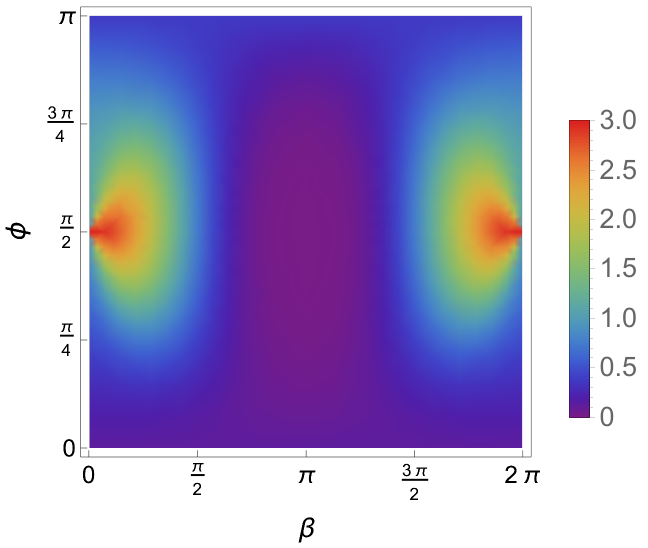}}
    \caption{Spin correction to the differential cross sections of graviton photoproduction in the ``++'' channel is polar asymmetric due to the spin. The ``-- --'' channel is related to the ``++'' channel by Eq.~(\ref{eq: xsec_symmetry}). For graviton-spin alignment/anti-alignment, the spin dependence at $\mathcal{O}(S^2)$ vanishes, but is present at $\mathcal{O}(S)$. Here we present the spin corrections for (1)$\alpha=0, \theta=0$, (2)$\alpha=0, \theta=\pi/3$, (3)$\alpha=0, \theta=\pi/2$.}
    \label{fig:conserving}
\end{figure*}

\begin{figure*}[h]
    \centering
    \subfigure[$\alpha=0, \theta=0$]{\includegraphics[width=0.32\columnwidth]{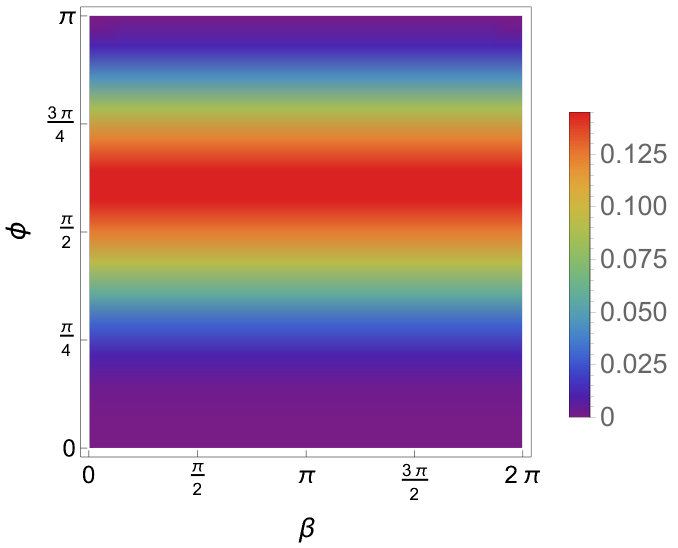}}
    \subfigure[$\alpha=0, \theta=\pi/3$]{\includegraphics[width=0.315\columnwidth]{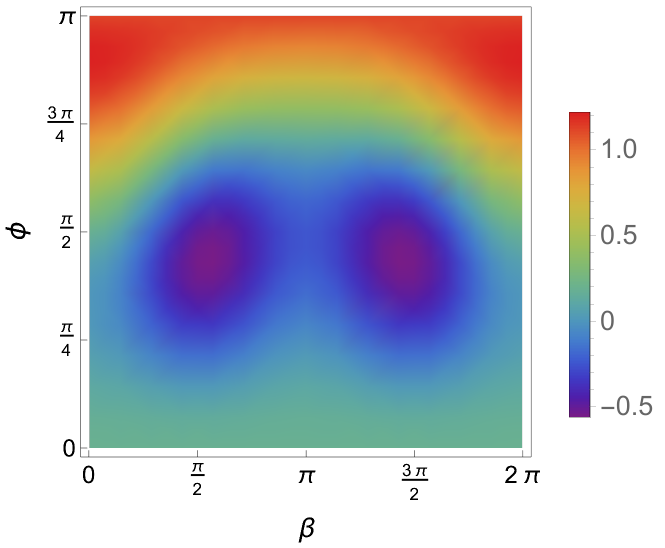}} \subfigure[$\alpha=0, \theta=\pi/2$]{\includegraphics[width=0.32\columnwidth]{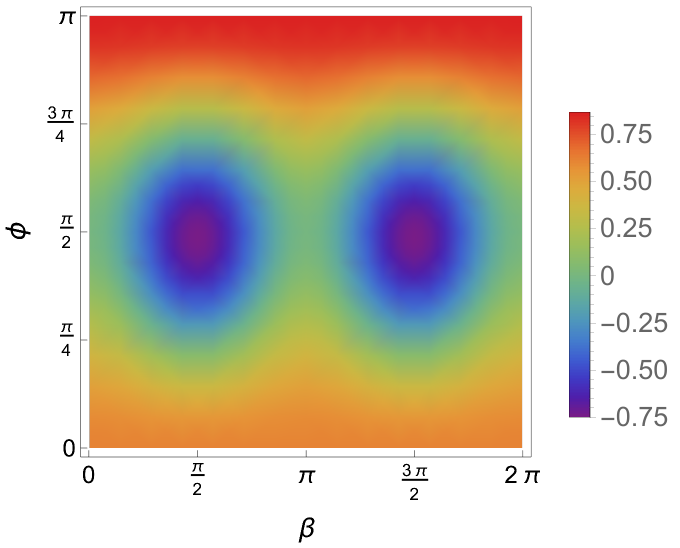}}
    \caption{Spin correction to the differential cross sections of graviton photoproduction in the ``+--'' channel is also polar asymmetric due to the spin, as illustrated in (c). For graviton-spin alignment/anti-alignment, the spin dependence at $\mathcal{O}(S^2)$ vanishes, but is present at $\mathcal{O}(S)$. Here we present the spin corrections for (a)$\alpha=0, \theta=0$, (b)$\alpha=0, \theta=\pi/3$, (c)$\alpha=0, \theta=\pi/2$.}
    \label{fig:flipping}
\end{figure*}

In Fig.~\ref{fig:conserving} and Fig.~\ref{fig:flipping}, we show representative examples of differential cross sections in the helicity-conserving ``++'' channel and the helicity-reversing ``+--'' channel. The geometric asymmetry in the polar angular dependence due to the spin is visible in these individual helicity channels. When the incident graviton momentum is aligned or anti-aligned with the spin, the $\mathcal{O}(S^2)$ contribution drops, leaving only the $\mathcal{O}(S)$ correction as shown in the sub-figure (a)'s of Fig.~\ref{fig:conserving} and Fig.~\ref{fig:flipping}.

\twocolumngrid

% --- Bibliography ---
\bibliographystyle{apsrev4-2}
\bibliography{refs}

\end{document}